\newcommand{\eqnref}[1]{Eq.\ (\ref{#1})}
\newcommand{\Eqnref}[1]{Equation (\ref{#1})}

\newcommand{\eqnarefs}[2]{Eqs.\ (\ref{#1}) and (\ref{#2})}
\newcommand{\Eqnarefs}[2]{Equations (\ref{#1}) and (\ref{#2})}
 
\newcommand{\step}[1]{step ($#1$)}
\newcommand{\steps}[2]{steps ($#1$) and ($#2$)}
\newcommand{\stepd}[2]{steps ($#1$) -- ($#2$)}

\newcommand{\Steps}[2]{Steps ($#1$) and ($#2$)}
\newcommand{\secref}[1]{Sec.\ \ref{#1}}
\newcommand{\secsref}[2]{Secs.\ \ref{#1} and \ref{#2}}
\newcommand{\sectref}[3]{Secs.\ \ref{#1}, \ref{#2}, and \ref{#3}}

\newcommand{\figref}[1]{Fig.\ \ref{#1}}
\newcommand{\figsref}[2]{Figs.\ \ref{#1} and \ref{#2}}

\newcommand{\subfigref}[2]{Fig.\ \ref{#1}(#2)} 
\newcommand{\subFigref}[2]{Figure \ref{#1}(#2)} 
\newcommand{\Figref}[1]{Figure \ref{#1}}
\newcommand{\Figsref}[2]{Figures \ref{#1} and \ref{#2}}
\newcommand{\bigO}[1]{$O(#1)$}
\newcommand{\ErdosRenyi}[0]{Erd\H os-R\'enyi}
\newcommand{\ermodel}[0]{Erd\H os-R\'enyi null model}
\newcommand{\wbar}[0]{\overline{w}}
\newcommand{\kpowavg}[0]{\langle k\rangle_\alpha}

\newcommand{\Ham}{\mathcal{H}}
\newcommand{\etal}{\emph{et al}.}

\newcommand{\ie}{\emph{i.e.}}
\newcommand{\myfig}[5]{%
\begin{figure}[{#5}]%
\includegraphics[keepaspectratio,width=#4,angle=0]{#1}{\caption{#2}\label{#3}}%
\end{figure}}

\newcommand{\gammarb}[0]{\gamma_{{\protect \phantom{}_{RB}}}}
\newcommand{\gammaer}[0]{\gamma_{{\protect \phantom{}_{ER}}}}

\documentclass[aps,pre,twocolumn,showpacs,amsmath,amssymb]{revtex4}
\usepackage{graphicx,dcolumn,bm,verbatim,color,hyperref}

\begin{document} 

\bibliographystyle{apsrev} 

\title{Local resolution-limit-free Potts model for community detection}

\author{Peter Ronhovde}
\author{Zohar Nussinov}
\affiliation{Department of Physics, Washington University in St. Louis, 
Campus Box 1105, 1 Brookings Drive, St. Louis, Missouri 63130, USA}%

\date{\today}

\begin{abstract}
We report on an exceptionally accurate spin-glass-type Potts model 
for community detection.
With a simple algorithm, we find that our approach is at least 
as accurate as the best currently available algorithms and robust 
to the effects of noise.  
It is also competitive with the best currently available algorithms 
in terms of speed and size of solvable systems.
We find that the computational demand often exhibits superlinear scaling 
\bigO{L^{1.3}} where $L$ is the number of edges in the system, 
and we have applied the algorithm to synthetic systems as large as 
$40\times 10^6$ nodes and over $1\times 10^9$ edges.
A previous stumbling block encountered by popular community detection 
methods is the so-called ``resolution limit.''
Being a ``local'' measure of community structure, our Potts model is 
free from this resolution-limit effect, and it further remains a local 
measure on weighted and directed graphs.
We also address the mitigation of resolution-limit effects for two other 
popular Potts models.
\end{abstract}


\maketitle{}

\vskip 0.1in

\section{Introduction} \label{sec:introduction}

``Community detection'' describes the problem of finding closely 
related sub-groups within a network. 
Applications of the problem are extremely broad finding realizations 
in the World Wide Web \cite{ref:danonhetero,ref:lanc}, 
food webs \cite{ref:gnsocbio}, 
social networks \cite{ref:gnsocbio}, 
protein interactions \cite{ref:palla}, 
consumer purchasing patterns \cite{ref:clausetlocal}, 
mobile phone networks \cite{ref:blondel}, 
criminal networks \cite{ref:xuchen},
epidemiology \cite{ref:masudaimmune},
biological networks \cite{ref:gnsocbio,ref:guimera},
and other areas. 
A recent review of community detection appears 
in Ref.\ \cite{ref:fortunatophysrep}.

In this paper, we present an improvement to the Potts model 
as applied to community detection, and we demonstrate that it 
is extremely accurate, robust to noise, and competitive with 
the best available methods in terms of computational speed 
and the size of solvable systems. 
Our approach also corrects known problems encountered by other 
popular measures that do not properly resolve small communities  
when large sets of data are examined.

The data in networks can often be cast as a graph consisting 
of members represented by nodes with pair-wise relationships 
between the nodes represented by edges.
In general, these relationships can be specified in one 
direction along an edge, 
and they can be weighted or unweighted.
The goal of community detection is to efficiently separate 
clusters of closely related nodes from each other. 
Each cluster will have a proportionally higher number 
of internal edges compared its external connections 
to each other community in the partition.

The most popular quantitative community measure is that 
of ``modularity'' which was originally introduced by 
Newman and Girvan \cite{ref:gn}.
This measure constituted a work that transformed the field.
Modularity measures the deviation of a proposed community 
structure compared to what is expected from an ``average'' 
case based on a particular random distribution 
(a ``null model'').

Our approach is a physics-inspired method that casts community 
detection as a Potts model spin glass.
Communities correspond to Potts model spin states, and the 
associated system energy indicates the quality of a candidate 
partition.
Some earlier approaches utilizing Potts models are 
in \cite{ref:blatt} and \cite{ref:reichardt}.
Our particular model was originally inspired by a 
minimal cut method by Djidjev \cite{ref:djidjev} 
which is equivalent to modularity.
The resulting generalized Hamiltonian was previously 
presented by Reichardt and Bornholdt (RB) \cite{ref:smcd}.
In their specific implementation, 
RB generalized null model based approaches to community 
detection, including modularity as a special case, 
and elaborated on the connection between physics 
and community detection.
Two other Potts model approaches are by Hastings \cite{ref:hastings}
and Ispolatov \etal{} \cite{ref:ispolatov}.

Modularity and the RB Potts model (RBPM) utilize a random null model 
selected to evaluate the strength of a proposed community partition. 
Larger deviations 
(more intra-community links and fewer inter-community links) 
from the random case indicate better community structure. 
A null model is usually based on parameters of the graph being 
examined which allows the measure to ``scale'' to arbitrary 
graphs in an objective manner (see \secref{sec:localglobal}).
Typical null models for the RBPM are: 
($i$) an \ErdosRenyi{} null model (RBER) in which all edges are 
equally likely to be connected and 
($ii$) the configuration null model (RBCM) in which edge connection 
probabilities are based on the current graph's degree distribution.
For modularity, the dependence on the null model is inherent 
to the definition of the measure.
Within the RB scheme, the dependence on a null model is 
introduced by design.

Fortunato and Barth\'elemy \cite{ref:fortunato} later determined
that modularity optimization can result in incorrect community 
divisions due to a \emph{resolution limit}.
The resolution limit is an inherent scaling in the expected 
number of communities $q$ which roughly scales as $\sqrt{L}$  
where $L$ is the total number of edges in the graph.
The RBPM model is also subject to a resolution limit 
\cite{ref:kumpulaResLim} due to how it is cast by design, 
analogous to modularity, in terms of an arbitrary null 
model comparison.
The number of communities roughly scales as $\sqrt{\gammarb L}$, 
where $\gammarb$ is a weight applied to the null model comparison. 
Optimizing either measure 
(maximizing modularity or minimizing the Potts model energy) 
tends to merge small clusters in large systems, 
or it may incorrectly partition large communities.
Although the RBPM allows for an arbitrary choice 
of null model, the resolution limit was shown to persist 
\cite{ref:kumpulaResLim} regardless of the null model 
that is used.  

Our approach avoids a null model comparison \cite{ref:rzone}. 
Instead, it penalizes for missing edges directly in the energy 
sum \cite{ref:hastingsnote}.
In effect, a community is defined by its edge density as opposed 
to allowing each graph to independently define a community through 
the use of a null model.
One consequence is that it removes the ability of the model to 
automatically scale the solution based on global properties of a 
graph (see \secref{sec:localglobal}), but the change results in a 
robust model with significant improvements to several desirable 
properties.

In this paper, we demonstrate a simple but effective implementation 
of the $q$-state Potts model to community detection.
In \secref{sec:hamiltonians}, we discuss our Potts model and some 
of its properties along with the RBPM and its main variants.
We also explain the concept of the \emph{resolution} of a partition.
In \secref{sec:algorithm}, we present our algorithm, and 
\secref{sec:GNtest} illustrates its accuracy compared to several 
other approaches.
Our Potts model and the RBCM model are directly compared 
in \secref{sec:APMvsRB}.
Issues regarding local and global measures and the resolution limit 
for general graphs are addressed in \secref{sec:resolutionlimit}.
We solve two examples in \secref{sec:examples} and conclude 
in \secref{sec:conclusion}.
Appendix A argues that the \emph{unweighted} variant of the 
RBER model can be strengthened to eliminate the resolution limit. 
In Appendix B, we explain the variation of information (VI) metric 
which we use in \secref{sec:APMvsRB}.
Appendices C and D elaborate on details related to the benchmark
in \secref{sec:noisetest}.
Appendix E explains a \emph{community detection transition}
observed in the same benchmark. 

\section{Potts Model Hamiltonians} \label{sec:hamiltonians}

Quality functions in community detection evaluate the ``best'' 
partitions based on at least two criteria:
Edges inside a community strengthen the community.
In order to consistently avoid a trivial solution (a single 
community), a quality function must also apply a ``penalty 
function.''
The most common method compares community degree distributions 
to an ``expected'' distribution based on a null model, but other 
approaches are possible.

\subsection{Absolute Potts model}

Our Potts model directly penalizes for missing edges within a 
community.  
The result is a robust model that is highly accurate, a local 
model for general graphs (weighted, unweighted, and directed), 
and \emph{free of the resolution limit}.
We also connect the introduced model weight $\gamma$
to the \emph{resolution} of a system and relate the interaction 
energies to the stability of communities.

\subsubsection{Hamiltonian}

We construct the Potts model with the following considerations.
Edges \emph{inside} communities and missing edges \emph{outside} 
communities are both favorable for a well-defined community structure, 
so the energy of the system is lowered by these arrangements. 
The opposite holds for edges outside communities and missing edges 
inside communities. 
This generalized Potts Hamiltonian is 
\cite{ref:smcd}
\begin{equation}
	\Ham(\{ \sigma \}) = - \frac{1}{2} \sum_{i\neq j}
	{\big( a_{ij} A_{ij} - b_{ij} J_{ij} \big)  
	 \big[ 2\delta (\sigma_i,\sigma_j) - 1 \big] } 
	\label{eq:startingH}
\end{equation}
where $\{A_{ij}\}$ is the set of adjacency matrix elements: 
$A_{ij}= 1$ if nodes $i$ and $j$ are connected and is $0$ 
if they are unconnected, and $J_{ij} \equiv (1-A_{ij})$.
The edge weights ($\{a_{ij}\}$ and $\{b_{ij}\}$)
and connection matrices ($\{A_{ij}\}$ and $\{J_{ij}\}$) 
are defined by the system. 
The Potts spin variable $\sigma_{i}$ takes an integer 
value in the range $1 \le \sigma_{i} \le q$ 
which designates the community membership of node $i$
(node $i$ is in community $k$ if $\sigma_{i}=k$).
The number of communities $q$ can be set as a constraint,
or it can be determined from the lowest energy configuration.
The Kroneker delta $\delta (\sigma_i,\sigma_j) = 1$ if 
$\sigma_i = \sigma_j$ and $0$ if $\sigma_{i} \neq \sigma_{j}$.

The spin glass type Potts model of \eqnref{eq:startingH} can be reduced, 
up to an additive constant,
to a form that greatly simplifies implementation.
For comparison, we introduce a form similar in appearance 
to the notation used by RB 
\begin{equation}  \label{eq:ourmodel}
	\Ham(\{ \sigma \} ) = -\frac{1}{2} \sum_{i\neq j}
	    \big( a_{ij} A_{ij} - \gamma b_{ij} J_{ij} \big) 
	    \delta ( \sigma_i,\sigma_j ).
\end{equation}
Spins interact only with other spins 
in the same community  ($\sigma_{i} = \sigma_{j}$).
The introduced factor of $\gamma$ allows the model to scale 
the relative strengths of the connected and missing edge weights.
The generality of the weights ($\{a_{ij}\}$ and $\{b_{ij}\}$) 
\cite{ref:rzone,ref:rzmultires} enables the study of directed graphs, 
weighted graphs, and graphs with missing link weights 
(\ie{}, levels of ``repulsion''). 
Traag and Bruggeman \cite{ref:traagPRE} also presented a generalization 
of the RBCM that similarly allows for ``negative'' link weights.
Unweighted graphs use edge weights of $a_{ij}=b_{ij}=1$.

The Hamiltonian of \eqnref{eq:ourmodel} describes a system wherein 
spins in the same community interact ferromagnetically if they are 
connected and antiferromagnetically if they are not connected.
We identify communities by minimizing \eqnref{eq:ourmodel},
and despite a global energy sum, 
our model is a \emph{local} measure of community structure 
(see \secref{sec:resolutionlimit}).
We refer to \eqnref{eq:ourmodel} as an \emph{``absolute'' Potts model} 
(APM) as it is not defined relative to a null model. 
Although our analysis in this paper will focus on the static APM, 
it is defined for both {\em static} systems and \emph{dynamic} 
networks with time-dependent weights and adjacency matrices.

\subsubsection{Resolution} \label{sec:resolutiondef}

Intuitively, the \emph{resolution} of a community partition 
is set, on average, by the strength of intra-community
connections. That is, the resolution of the partition may be specified by the 
typical edge density of the communities within the partition. 
Communities with substantially different edge densities 
have different qualitative features.

In social networks for example, a partition intending to convey 
the ``close friends'' within a network would intuitively have a 
higher typical edge density than a partition that includes all 
``acquaintances'' since the disparate acquaintances are much
less likely to know each other.
Ideally, a partition should contain communities that convey 
similar qualitative information 
(\ie{}, similar ``levels'' of association). 
In practice, it will contain communities with different 
edge densities, but intuitively the differences would not be drastic 
for a given resolution.

For unweighted graphs, the edge density $p_s$ of community $s$ 
is $p_s = \ell_s/\ell_s^\mathrm{max}$ where $\ell_s$ is the number 
of edges in the community.
$\ell_s^\mathrm{max}=n_s(n_s-1)/2$ where $n_s$ is the number of nodes.
The model weight $\gamma$ in \eqnref{eq:ourmodel} 
is related to the \emph{minimum} edge density of each community,
\begin{equation} \label{eq:gammadensity}
  p_\mathrm{min}\ge \frac{\gamma}{\gamma+1},
\end{equation}
which is determined by calculating the minimum community density 
that gives an energy of zero or less.
Alternately, we can use an inductive argument based on the maximum 
intercommunity edge density that causes two arbitrary communities 
to merge. 
For weighted graphs, we define a ``weight density'' 
$p_s\equiv w_s/w_s^\mathrm{max}$ where $w_s$ is the sum of all weighted 
edges in community $s$ and the ``maximum weight'' 
$w_s^\mathrm{max}\equiv \overline{w}_s \ell_s^\mathrm{max}$
where $\overline{w}_s$ is the average edge weight.
The minimum density is $p_\mathrm{min}\ge \gamma/(\gamma+\overline{w}_s/\overline{u}_s)$
where $\overline{u}_s$ is the average weight of the missing links.
Without $\gamma$, the model is restricted to solving one particular 
resolution of a system. 
This relation between $\gamma$ and the community density is distinctly 
different from a resolution limit because the communities are determined 
through only \emph{local} constraints (see \secref{sec:resolutionlimit}).

\subsubsection{Community and node stability}

From \eqnref{eq:ourmodel}, the interaction energy $E_{rs}$ 
between communities $r$ and $s$ is 
\begin{equation} \label{eq:nodeErs}
  E_{rs} = -w_{rs} + \gamma u_{rs}
\end{equation}
where $w_{rs}$ is the energy sum over all edges and $u_{rs}$ 
is the energy sum over all \emph{missing} links strictly 
\emph{between} the two communities.
$E_{ss}\equiv E_s$ is the internal energy of community $s$ where 
the energy sum is over all \emph{internal} edges and missing links.
When $E_s\simeq 0$, the assignment of community $s$ is more 
sensitive to local perturbations.

Similarly, the interaction energy $E_{ri}$ of node $i$ 
with community $r$ is given by \eqnref{eq:nodeErs}.
If $E_{si} - E_{ri} \simeq 0$ for node $i$ in community $s$,
then the node is susceptible to displacement by system perturbations.
When a node contributes a large fraction of the energy $E_s$ of its 
own community, the community is susceptible to disruption if the node 
is moved.
\Eqnref{eq:nodeErs} indicates the strong local behavior 
of the APM (see \secref{sec:resolutionlimit}). 
For general graphs, the interaction energy of node $i$ or 
community $s$ is measured \emph{only} by its \emph{own} edges 
or missing links with each community.

\subsection{RB Potts models}
\label{sec:RBmodels}

We compare the APM to the RBPM in order to demonstrate 
improvements in accuracy and locality despite the apparent 
similarity in the models.
The RBPM, using an arbitrary null model, is defined as 
\cite{ref:smcd}
\begin{equation}
	\Ham_{RB}(\{ \sigma \} ) = 
	  -\frac{1}{2} \sum_{i\neq j}{ \left(  A_{ij} - \gammarb p_{ij}\right)
	        \delta (\sigma_i,\sigma_j) }
	\label{eq:RBmodel}
\end{equation}
where we include the overcounting scale factor of $1/2$.
The term $p_{ij}$ is the probability that nodes $i$ and $j$ 
are connected, and it incorporates the dependence 
on the arbitrary null model.
$\gammarb$ is the weight applied to the null model.
The most frequently used null models are an \ermodel{} 
and the configuration null model (see \secref{sec:introduction}).
For later reference, they are explicitly given 
by $p_{ij}=p$ for the \ermodel{}
\begin{equation}
	\Ham_{RB}^{ER}(\{ \sigma \} ) = 
	  -\frac{1}{2} \sum_{i\neq j}{
	    \left(  A_{ij} - \gammarb p\right)
	  \delta (\sigma_i,\sigma_j) }
	\label{eq:RBmodelER}
\end{equation}
and by $p_{ij}={k_i k_j}/({2L})$ for the configuration null model
\begin{equation}
	\Ham_{RB}^{CM}(\{ \sigma \} ) = 
	  -\frac{1}{2} \sum_{i\neq j}{
	    \left(  A_{ij} - \gammarb \frac{k_i k_j}{2L} \right)
	  \delta (\sigma_i,\sigma_j) },
	\label{eq:RBmodelCM}
\end{equation}
where 
$k_i$ is the degree of node $i$.
\Eqnref{eq:RBmodelCM} appears to be the more preferred model since 
the configuration null model incorporates information about the 
degree distribution of the graph under consideration.

When $\gammarb = 1$, the RBCM of \eqnref{eq:RBmodelCM} is equivalent 
to modularity \cite{ref:smcd} up to a scale factor of $-1/L$.
The APM can be made equivalent to the RBER model for \emph{unweighted} 
graphs \cite{ref:RBERnoisenote} (see also Appendix A). 
We address weighted generalizations of both models and 
their effect on model locality in \secref{sec:weightedlocal}. 
Despite the similar forms of the Hamiltonians of 
\eqnarefs{eq:ourmodel}{eq:RBmodel}, the model weights $\gamma$ 
and $\gammarb$ perform distinctly different roles in the two models.  
In the APM, $\gamma$ directly adjusts the weight applied to \emph{missing edges}.
In the RBCM, $\gammarb$ adjusts the weight applied to the \emph{null model}.
We contrast the accuracy of the APM and the RBCM in \secref{sec:APMvsRB}.

\section{Algorithm} \label{sec:algorithm}

Our algorithm moves nodes by identifying which community they may be moved 
into so that the system energy is lowered.
The algorithm proceeds until no more node moves are possible. 
This ``orthogonal steepest descent'' algorithm (selecting the path of steepest 
descent for only one spin $\sigma_i$ at a time) is extremely fast. 
We introduced our initial implementation of the algorithm in \cite{ref:rzone}.
A summary of the efficiency of several algorithms appears in \cite{ref:danon}. 
A number of algorithms were compared in \cite{ref:noack} and \cite{ref:lancLFRcompare} 
where algorithms similar to ours performed very well when optimizing modularity.
Combined with the APM, it is exceptionally accurate. 
The steps of the algorithm are:

(1) \emph{Initialize the system}. 
Initialize the connection matrices ($A_{ij}$ and $J_{ij}$) 
and edge weights ($a_{ij}$ and $b_{ij}$).
The system begins in a ``symmetric'' state wherein each node forms 
its own individual community ($q_0=N$).
If the number of communities $q$ is constrained 
(e.g., \figsref{fig:accuracyplot}{fig:zacharyclub}), 
we randomly initialize the system into $q_0=q$ communities.  

(2) \emph{Optimize the node memberships}. 
Sequentially ``pick up'' each node and scan its neighbor list.
Calculate the energy change as if it were moved to each connected 
cluster.
Immediately place it in the community with the lowest energy 
(optionally allowing zero energy changes).
Each iteration through all nodes is \bigO{L}. 

(3) \emph{Iterate until convergence}. 
Repeat \step{2} until an energy minimum is reached where 
no node moves will further lower the system energy. 

(4) \emph{Test for a local energy minimum}. 
Manually merge any connected communities if the merge(s) will 
further lower the energy of the system. 
If any merges are found, return to \step{2} for any additional
node-level refinements.
We estimate that the computational cost is \bigO{L\log q}
which is generally smaller than the node optimization 
cost in \steps{2}{3}.
     
(5) \emph{Repeat for several trials}. 
Repeat \stepd{1}{4} for $t$ independent ``trials'' and select the 
lowest energy result as the best solution. 
By a trial, we refer to a copy of the network in which the 
initial system is randomized.

The symmetric initialization for the nodes in \step{1} is 
not uncommon in the 
literature \cite{ref:blondel,ref:clausetlarge,ref:LPA,ref:gudkov}.
\Steps{2}{3} are the fundamental elements of the algorithm which 
are similar to portions of algorithms used 
elsewhere \cite{ref:blondel,ref:LPA}.
The number of iterations is generally \bigO{10} for large systems,
but it can be higher for ``hard'' problems. 
In \step{4}, the community merge test is sometimes necessary 
because certain configurations, particularly heavily weighted 
graphs with $\gamma\ll 1$, more easily trap the node-level 
refinements [\steps{2}{3}] in local energy minima.
The merge test is not generally a major concern for $\gamma\ge 1$.

The order of node moves is significant, so the additional trials 
in \step{5} sample different regions of the energy landscape and 
can yield different solutions even with the symmetric initialization 
in \step{1}.
We optimize solutions by increasing the number of trials where the 
greatest benefit occurs for problems of ``intermediate'' difficulty
(e.g., see the data for the APM in \figref{fig:accuracyplot}).
The number of trials $t$ is generally \bigO{10} or less.

Empirically, the overall solution cost often scales as 
\bigO{tL^{1.3}\log k} where $k$ is the average node degree.
The factor of $\log k$ applies for large sparse matrix systems.
The algorithm can accurately scale to at least \bigO{10^7} nodes 
and \bigO{10^9} edges with a calculation time of several hours 
\cite{ref:computerusedone} (see \secref{sec:examples}).

\section{Accuracy Compared to Other Algorithms} 
\label{sec:GNtest}

\myfig{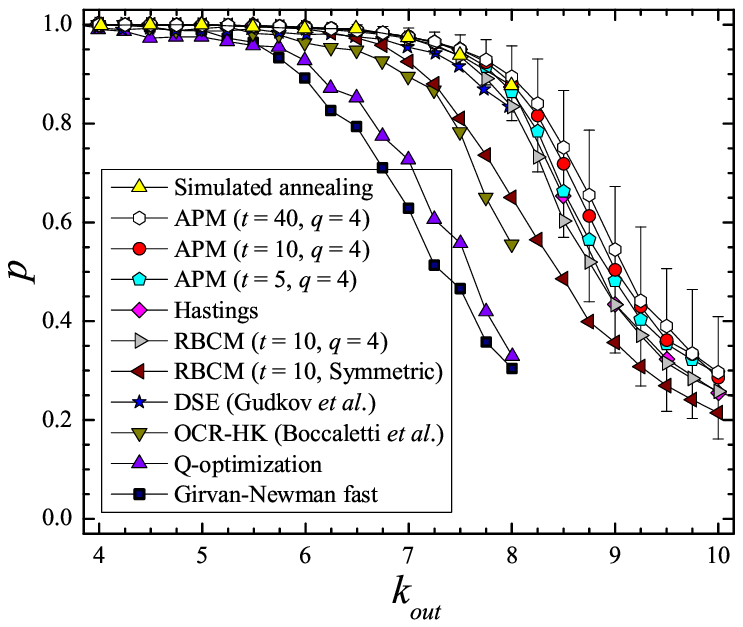}{(Color online) 
Plot of the percentage of correctly identified nodes $p$ versus 
the average external degree $k_{out}$ \cite{ref:newmanfast}. 
The average node degree is $k=16$.
The data for the APM of \eqnref{eq:ourmodel} and the RBCM 
of \eqnref{eq:RBmodelCM} both use $\gamma =\gammarb =1$.
Both models use the algorithm in \secref{sec:algorithm} with $q=4$ 
by constraint (see text regarding the RBCM/Symmetric data).
The APM is at least as accurate as SA (error bars are for $t=10$ 
optimization trials), and the RBCM performs excellently also.
Each point is an average over $500$ runs.
}{fig:accuracyplot}{0.95\linewidth}{t}

We test the accuracy of our method compared to several other 
algorithms using a common benchmark \cite{ref:newmanfast}.
The benchmark is very small by current standards with an 
unrealistically symmetric community structure, but its frequent 
use provides a means of comparing the accuracy of various algorithms 
that have been presented in the literature over time.
The problem defines a system of $N=128$ nodes in $q=4$ clusters 
of $n=32$ nodes each.
Each node is assigned an average of $k=16$ edges of which $k_{in}$ 
are randomly assigned inside its own community.
$k_{out}$ edges are randomly assigned to nodes in other communities 
such that $k=k_{in}+k_{out}$. 
We then attempt to verify the defined community structure.

In \figref{fig:accuracyplot}, we plot the ``percentage'' of correctly 
identified nodes $p$ as a function of $k_{out}$.
For consistency with other data in \figref{fig:accuracyplot}, we use 
the same measure of percentage accuracy as Newman \cite{ref:newmanfast}.
We use $q=4$ communities by constraint and test several levels of 
optimization ($t=5$, $10$, and $40$).
At $t=10$, our method maintains an accuracy rate of $95\%$ or better 
up to $k_{out}=7.5$.

Several sets of data 
were assimilated by Boccaletti \etal{} \cite{ref:boccaletti} 
where the most accurate algorithm was simulated annealing 
(SA) although it is computationally demanding \cite{ref:danon}.
Other accurate data are by Hastings \cite{ref:hastings}, Gudkov 
\etal{} \cite{ref:gudkov}, and our algorithm in \secref{sec:algorithm} 
applied to the RBCM with $\gammarb=1$ (modularity) and $t=10$.
Our algorithm is as accurate as SA when used with the APM.

The APM, one set of our data for the RBCM, and the data by Hastings 
impose $q=4$ as a constraint; so using a constrained $q$ may affect 
the accuracy rate in this problem.
The \emph{initial state} of the system substantially influences the 
accuracy of our algorithm for the RBCM  when $q$ is unconstrained 
and when starting from an initial state of one node per cluster 
(symmetric) or a random state (not depicted) \cite{ref:RBCMGNtestnote}.
A recent analysis \cite{ref:lancLFRcompare} showed our multiresolution 
algorithm \cite{ref:rzmultires} applied to this benchmark using the 
APM with unconstrained $q$ where it was also very accurate, among the 
best of tested algorithms.


\section{Accuracy Comparison of Potts Models} 
\label{sec:APMvsRB}

\myfig{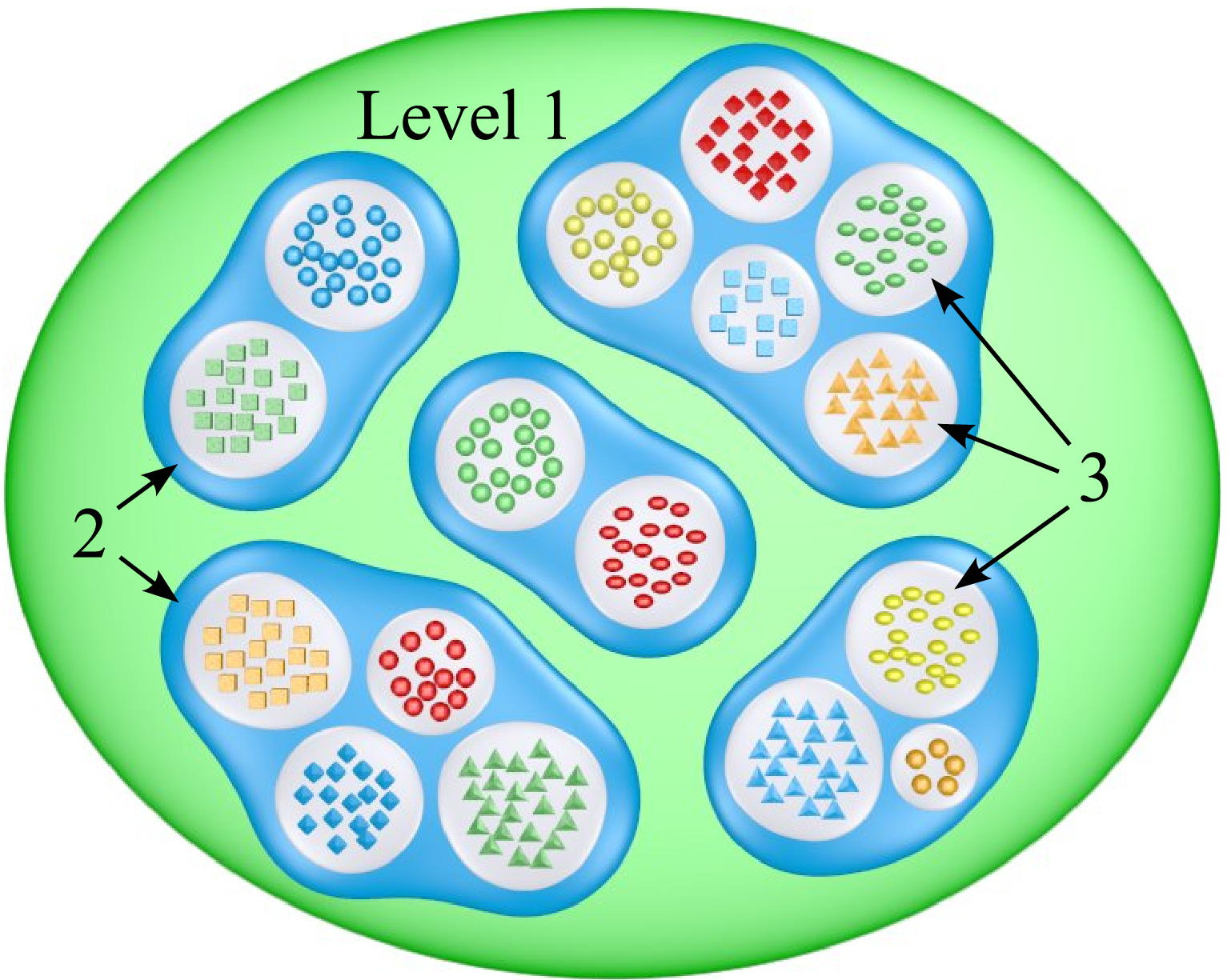}{(Color online) 
The figure depicts a simulated three-level heterogeneously-sized 
hierarchy with $N=256$ nodes \cite{ref:rzmultires,ref:datawebsite}.
The innermost level $3$ has $q_3=16$ communities with a randomly 
assigned average density of $\overline{p}_3=0.90$.
The intermediate level $2$ has $q_2=5$ communities and an average density 
of $\overline{p}_2=0.47$ that is constructed 
by connecting the constituent level $3$ sub-groups at an intercommunity 
edge density of $p_2=0.3$.
Level $1$ is the completely merged system with an average density 
of $\overline{p}_1=0.18$, and it is constructed by connecting nodes 
in different level $2$ communities with an intercommunity edge density 
of $p_1=0.1$.
}{fig:hierarchypic}{0.85\linewidth}{t}

\myfig{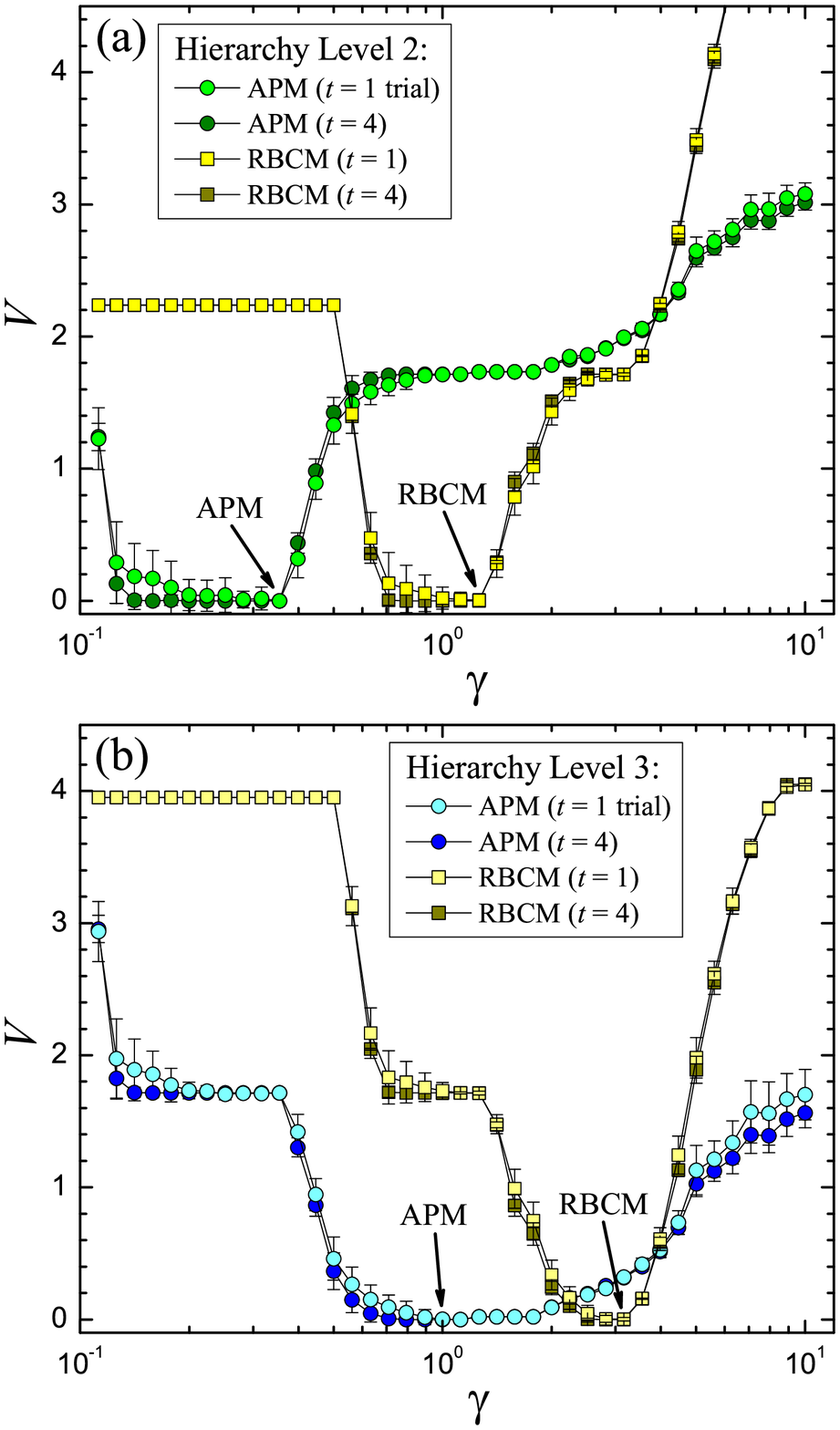}{(Color online) 
Plot of VI $V$ vs model weights $\gamma$ or $\gammarb$ for the APM 
of \eqnref{eq:ourmodel} and the RBCM (configuration null model)
of \eqnref{eq:RBmodelCM}, respectively. 
The plots illustrate how the model weights operate in the respective 
models.
We use the algorithm in \secref{sec:algorithm} for both models 
to identify the hierarchy depicted in \figref{fig:hierarchypic} 
using $t=1$ and $4$ trials.
We calculate VI with respect to level $2$ of the hierarchy in panel 
(a) and level $3$ in panel (b).
Both models exactly identify both levels of the hierarchy at $t=4$.
The APM perfectly identifies both levels at $t=1$ which is slightly 
better on average than the RBCM, and it has a more stable solution 
for level $3$.  Each point is an average over $100$ solutions.
}{fig:hierarchyplot}{0.95\linewidth}{t}

We compare the APM of \eqnref{eq:ourmodel} to the RBCM
of \eqnref{eq:RBmodelCM} with two test systems.
First, we solve for the different levels of the synthetic 
hierarchy depicted in \figref{fig:hierarchypic} with the 
results given in \figref{fig:hierarchyplot}.
Second, we create a set of strongly defined systems with high 
community edge densities and increasing levels of noise.
A sample graph is depicted in \figref{fig:powersamplegraph} 
with the results given in \figsref{fig:noisetest}{fig:noiseinittest}.
The APM proves to be very robust to noise in the system. 
We use the VI information metric $V$ (see Appendix B) to compare 
solved partitions with the constructed networks.

\subsection{Three-level hierarchy} 
\label{sec:hierarchy}

We identify two levels of a constructed hierarchy 
\cite{ref:datawebsite,ref:rzmultires} using both the APM 
and RBCM models.
The three-level hierarchy is depicted in \figref{fig:hierarchypic}, 
and the results are given in \figref{fig:hierarchyplot}.
The system has $N=256$ nodes divided into $q_3=16$ 
communities at level $3$ with sizes as noted in \figref{fig:hierarchypic}.
Edges in each community are randomly assigned with a probability 
of $\overline{p}_3=0.90$.
These communities are grouped as shown into $q_2 = 5$ communities 
that define level $2$ of the hierarchy.
The average internal density of level $2$ communities is 
$\overline{p}_2=0.47$ which are defined by randomly 
connecting nodes in the respective sub-groups of level $3$ 
at an intercommunity edge density of $p_2=0.3$.
Level $1$ is the completely merged system which is defined by randomly 
connecting nodes in sub-groups of level $2$ at an intercommunity 
edge density of $p_1=0.1$.

We apply the algorithm of \secref{sec:algorithm} to both models 
and solve a large range of model weights $\gamma$ or $\gammarb$, 
respectively, in order to illustrate the differences in the two models.
In \figref{fig:hierarchyplot}, we plot VI $V$ as a function $\gamma$ or 
$\gammarb$ \cite{ref:logscalenote} using $t=1$ and $4$ trials. 
VI is calculated between the respective solutions and the level $2$ or 
$3$ partitions of the hierarchy.
These data are then plotted in panels (a) and (b), respectively.
Both models exactly identify both levels of the hierarchy 
at $t=4$. 
The APM is slightly better in accurately identifying them with $t=1$, 
and it has a more ``stable'' solution in panel (b).

\subsection{Noise test} \label{sec:noisetest}

The concept of ``noise'' in community detection corresponds
to ``extra'' edges that connect a node to communities other than 
its best assignment(s).
In general, we cannot initially distinguish between edges 
contributing to noise and those constituting edges within 
communities of the best partition(s).
Community detection methods 
experience the effects of noise in at least two distinct ways:
($1$) The edges due to noise act to obscure the best partition(s) 
in an algorithm by creating ``confusion'' for early community 
assignments (a dynamical effect).
($2$) The extra edges influence the quantitative evaluation 
of the best community assignments (a ``metric'' effect).  
In some models, this second effect can negatively impact the 
contribution of edges that comprise the best communities.

\subsubsection{Benchmark}
\label{sec:benchmark}

\myfig{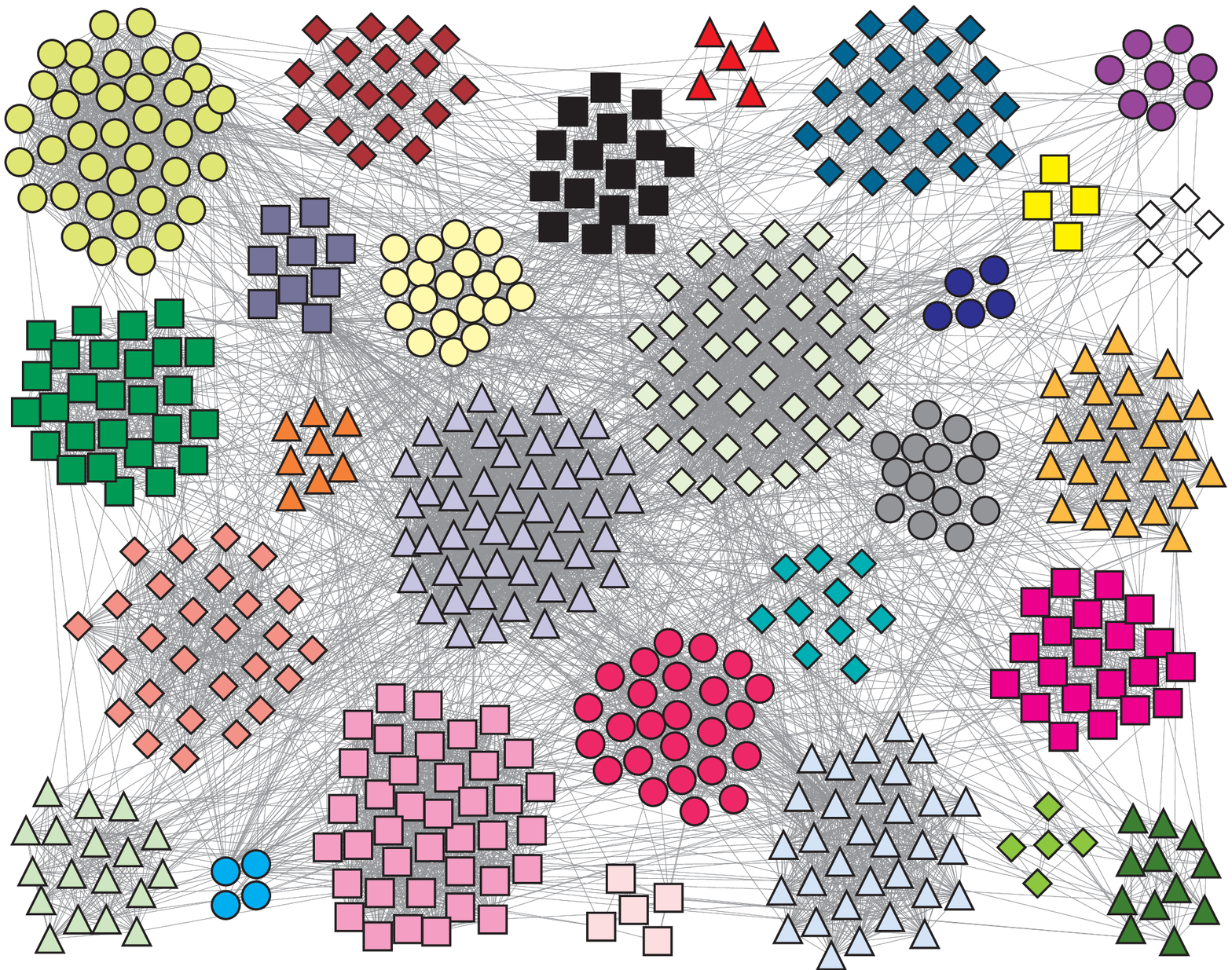}{(Color online) 
A sample graph with $N=512$ nodes for the noise test 
in \secref{sec:noisetest}.
In this sample, the node degrees are initially defined in a 
power-law distribution with an average $\kpowavg =5.4$, 
maximum $k_\mathrm{max}=100$, and exponent $\alpha =-2$.
Communities have a power-law distribution of sizes with a 
minimum $n_\mathrm{min}=4$, maximum $n_\mathrm{max}=50$, 
and exponent $\beta =-1$.
These communities are then strongly defined by connecting all 
internal community edges ($p_{in}=1$).
}{fig:powersamplegraph}{0.85\linewidth}{t}

\myfig{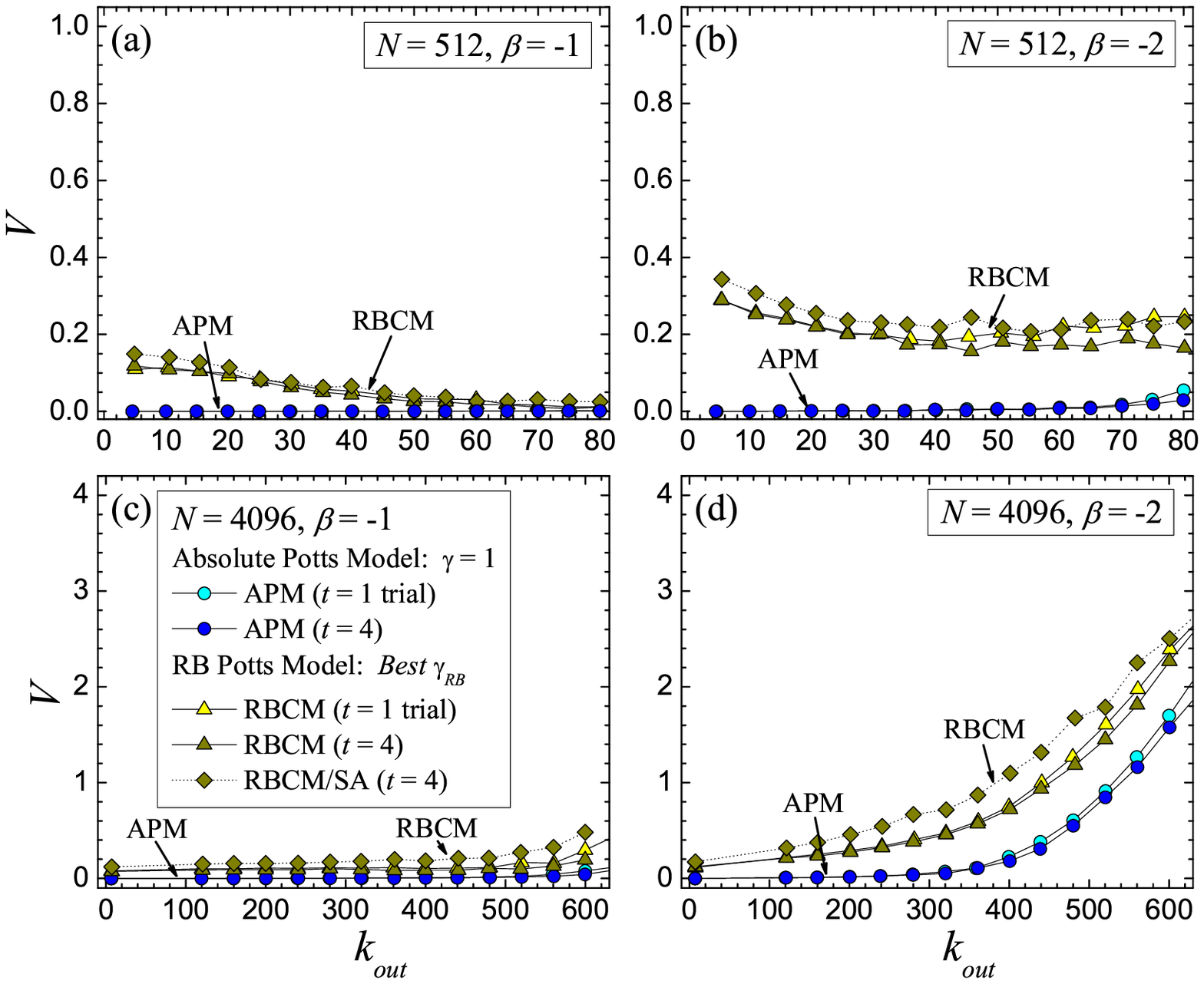}{(Color online) 
Plot of VI $V$ between solved and known test systems in
\secref{sec:noisetest} as a function of the average external 
node degree $k_{out}$.
The system sizes are  $N_{\mathrm{a},\mathrm{b}}=512$ in panels 
(a) and (b) 
(with a sample system depicted in \figref{fig:powersamplegraph})
and $N_{\mathrm{c},\mathrm{d}}=4096$ nodes in panels (c) and (d).
The graphs are solved with the APM 
and the RBCM 
using the algorithm in \secref{sec:algorithm} with $t=1$ and $4$ trials.
We use $\gamma =1$ for the APM on all solutions, and we subjectively
select the \emph{best} $\gammarb$ for the RBCM independently 
for each $k_{out}$ (see Appendix C).
For comparison, we also solve the system at this \emph{best} 
$\gammarb$ using SA.
System noise is randomly assigned in an approximate power-law 
degree distribution \cite{ref:powerinitnote} with an exponent 
$\alpha=-2$, an average degree $\kpowavg$, 
and maximum degrees of $k_{\mathrm{a},\mathrm{b}}^\mathrm{max}=100$ 
or $k_{\mathrm{c},\mathrm{d}}^\mathrm{max}=1000$, respectively.
Constructed communities are randomly assigned in a power-law size 
distribution \cite{ref:lancbenchmark} specified by an exponent
$\beta_{\mathrm{a},\mathrm{c}}=-1$ or $\beta_{\mathrm{b},\mathrm{d}}=-2$, 
minimum size $n_\mathrm{min}=4$, and maximum size $n_\mathrm{max}=50$.
Communities are then maximally connected with $p_{in}=1$.
Even with $t=1$, the APM is almost perfectly accurate for most 
tested parameters in this problem.
See Appendix E regarding the accuracy transitions in panel (d).
Data are averaged over $100$ graphs in panels (a) and (b)
and $25$ graphs in panels (c) and (d). 
}{fig:noisetest}{0.975\linewidth}{t}

\myfig{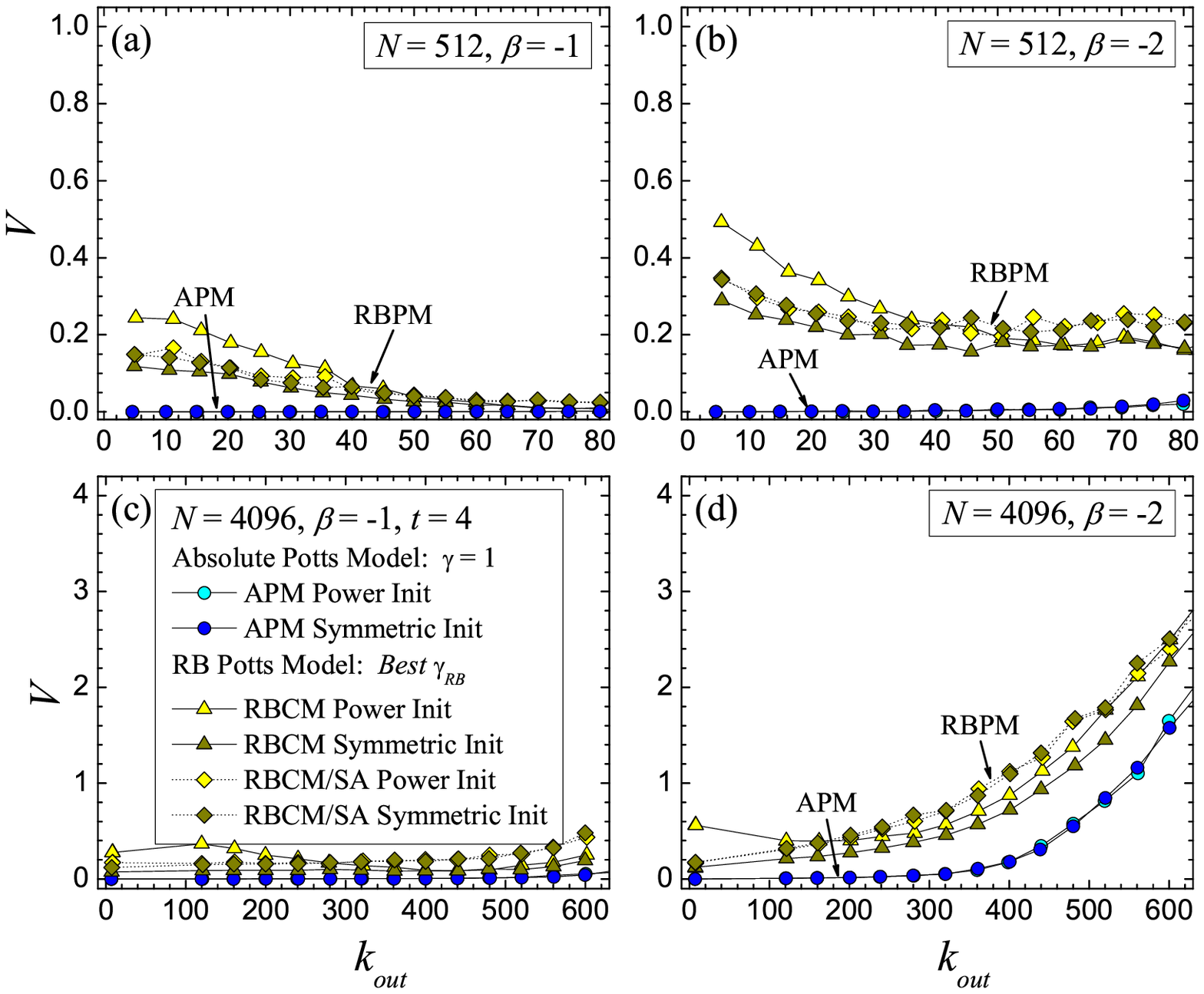}{(Color online) 
Plot of VI $V$ between solved and known test systems in
\secref{sec:noisetest} vs the average external node degree $k_{out}$.
In panels (a) through (d), system sizes are $N_{\mathrm{a},\mathrm{b}}=512$ 
and $N_{\mathrm{c},\mathrm{d}}=4096$ nodes, respectively.
The graphs are solved with the APM 
and the RBCM 
using the algorithm 
in \secref{sec:algorithm} with $t=4$ trials.
The constructed configurations are identical to those 
used in \figref{fig:noisetest}. 
In this plot, we test two different initial states for the 
solutions: a symmetric initial state 
and a random power-law distribution (see text).
We use $\gamma=1$ for the APM on all solutions, 
and we subjectively choose the \emph{best} $\gammarb$ for 
the RBCM for each $k_{out}$ (see Appendix C).
For comparison, the results for SA with $t=4$ are also 
depicted and are solved using this best value of $\gammarb$.
The APM and SA with the RBCM show no difference in accuracy between 
the symmetric and random initial states. 
A symmetric state appears to be the favored starting configuration 
for the RBCM when using a greedy algorithm in this benchmark.
In fact, this symmetric initial state allows the RBCM to slightly 
\emph{outperform} SA in accuracy (see text).
We average over $100$ graphs in panels (a) and (b) and $25$ graphs 
in panels (c) and (d).
}{fig:noiseinittest}{0.975\linewidth}{t}

We test the accuracy of APM and RBCM models with high levels 
of noise in a series of strongly defined systems with ``realistic'' 
distributions of community sizes.
Specifically, we define a set of communities with a power-law 
distribution of community sizes specified by an exponent $\beta$, 
minimum size $n_\mathrm{min}$, and maximum size $n_\mathrm{max}$.
We add random edges to the system, largely defining the intercommunity 
noise, based on a power-law distribution of node degrees given 
by an exponent $\alpha$, average power-law degree $\kpowavg$ 
(or minimum degree $k_\mathrm{min}$), 
and maximum degree $k_\mathrm{max}$ \cite{ref:powerinitnote}.
This initial framework is similar to a benchmark by Lancichinetti 
\etal{} \cite{ref:lancbenchmark,ref:lancLFRcompare}.
We then connect internal community edges at a high density $p_{in}$.

The strongly defined communities provide unambiguous partitions 
where the large level of noise does not significantly alter the 
optimal solutions (see \secref{sec:noisetolerance}).
This density-based definition of community structure is consistent 
with concepts proposed for community identification by Palla \etal{} 
\cite{ref:palla}.
We solve for the systems using the algorithm in \secref{sec:algorithm} 
for both models with $t=1$ and $4$ trials and using SA for the RBCM 
with $t=4$.

\subsubsection{Accuracy results}

\Figsref{fig:powersamplegraph}{fig:noisetest} show a sample system
and the first test results, respectively.
We use two system sizes of $N_{\mathrm{a},\mathrm{b}}=512$ 
and $N_{\mathrm{c},\mathrm{d}}=4096$ nodes, respectively.
The initial power-law degree distribution uses $\alpha =-2$; and
the maximum degree constraints are $k_{\mathrm{a},\mathrm{b}}^\mathrm{max}=100$ 
and $k_{\mathrm{c},\mathrm{d}}^\mathrm{max}=1000$, respectively.
We increment the average power-law degree $\kpowavg$ to vary 
the system noise (the average external degree $k_{out}\simeq\kpowavg$ 
for large systems).
Community sizes range from $n_\mathrm{min}=4$ to $n_\mathrm{max}=50$ 
nodes and are distributed according to $\beta_{\mathrm{a},\mathrm{c}}=-1$ 
or $\beta_{\mathrm{b},\mathrm{d}}=-2$, respectively.
The internal community edges are maximally connected at a density 
of $p_{in}=1$. 

We plot VI $V$ versus $k_{out}$ for both Potts models where VI is 
calculated between the solved partition and the generated graph
($V_{\mathrm{a},\mathrm{b}}^\mathrm{max} = 9$ and 
$V_{\mathrm{c},\mathrm{d}}^\mathrm{max} = 12$).
For the APM, we use $\gamma=1$ for every solution, and we allow zero 
energy moves after the system reaches an initially converged state. 
For the RBCM, we subjectively select the \emph{best} solution 
corresponding to the \emph{highest accuracy} $\gammarb$ independently 
determined for each $k_{out}$ given the \emph{known} answer 
(see Appendix C).
We further solve the system via SA at this best value of $\gammarb$
for comparison.
We average over $100$ graphs for each point in panels (a) and (b)
and $25$ graphs in panels (c) and (d).

In panels (a) and (c), the advantage in accuracy for the APM is 
modest. 
The accuracy of the RBCM increases in panels (a) and (b) at higher 
levels of noise due in part to the fact that the degree distribution 
is becoming more uniform as we increase $\kpowavg$ but keep
$k_\mathrm{max}$ constant.
While the RBCM performs excellently in many cases, the APM outperforms 
it to varying degrees for most tested parameters and levels of noise.
Moreover, the APM is often able to almost perfectly solve the system. 

The rapid increases in VI for both models in \subfigref{fig:noisetest}{d} 
are due to transition effects described in Appendix E.
We subjectively select the best $\gammarb$ in this paper, but 
note also that our algorithm can slightly outperform SA in accuracy 
in many cases \emph{for either Potts model} 
(see \secref{sec:initialconditions} and Appendix D). 
See \secref{sec:mitigatedRL} regarding how the high levels of noise 
in this test actually mitigate the effect of the resolution limit 
for the RBCM.

\subsubsection{Dependence on initial condition and SA accuracy}
\label{sec:initialconditions}

A community detection algorithm should ideally be robust with 
respect to the initial state that is used to solve the system.
We show that APM displays this feature, and we contrast the result 
with the RBCM when using the greedy algorithm in \secref{sec:algorithm}.
We further elaborate on the accuracy of SA compared to this 
greedy algorithm.

In \figref{fig:noiseinittest}, we plot VI $V$ vs $k_{out}$ where 
increasing $k_{out}$ corresponds to higher levels of system noise. 
We measure $V$ between the solved and constructed systems
where the defined systems are identical to the previous subsection.
We test both models beginning from two different initial states:
a symmetric initial state of one node per cluster with $q_0=N$ 
and a random power-law configuration with $q_0\simeq q$
which is different than the defined answer.
For simplicity, this random initial state uses the same distribution
parameters ($\beta_{\mathrm{a},\mathrm{c}}=-1$ or 
$\beta_{\mathrm{b},\mathrm{d}}=-2$, $n_\mathrm{min}=4$, 
and $n_\mathrm{max}=50$) that are used to generate the answers.

The best solutions for the APM are robust to the initial state 
of the system in this benchmark, including during the major accuracy 
transition in panel (d), despite using a greedy algorithm.
The symmetric initial state performs very well for the RBCM and is the 
favored starting configuration compared to the random power-law state. 
The situation is reversed for the RBCM on the benchmark in \figref{fig:accuracyplot}
in \secref{sec:GNtest} where the symmetric initialization performs worse 
than a random initial state with $q_0=4$ communities (with or without 
constraining $q$ in the dynamics \cite{ref:RBCMGNtestnote}) although 
that benchmark has an unrealistically symmetric community structure.
While optimizing the RBCM often provides excellent partitions, 
this difference in accuracy between initial states indicates that 
it is more easily trapped in unfavorable regions of the energy 
landscape than the APM when using a greedy algorithm. 

As expected, SA with the RBCM shows no difference in accuracy 
for either initial state, but the \emph{greedy} algorithm 
\emph{outperforms} SA in terms of accuracy when using a symmetric 
initial state (see also Appendix D).
This reduced accuracy for SA compared to a greedy algorithm is 
not isolated to this benchmark.
Lancichinetti and Fortunato \cite{ref:lancLFRcompare} compared 
the accuracy of several algorithms using their benchmark 
\cite{ref:lancbenchmark}.
One result in \cite{ref:lancLFRcompare} showed that a similar greedy 
algorithm optimizing modularity [equivalent to $\gammarb =1$
in \eqnref{eq:RBmodelCM}] by Blondel \etal{} \cite{ref:blondel} 
also outperformed SA in accuracy on that benchmark.
See also Good \etal{} \cite{ref:goodMC} regarding difficulties 
associated with modularity optimization in practical problems.

\subsubsection{Noise tolerance discussion}
\label{sec:noisetolerance}

Even at low levels of noise, these benchmark graphs exceed the 
proposed definition of so-called ``weak'' communities 
\cite{ref:radicchi}, but the communities are not ill-defined 
from an intuitive standpoint within the tested range of noise.
In panel (d) for example, at $k_{out}\simeq 370$ the average number 
of edges connecting a given node to another community is $\ell\simeq 1$ 
because the $k_{out}$ edges are randomly spread over 
$(q_\mathrm{b}-1)\simeq 370$ communities. 
This value is small compared to the average internal degree 
$k_{in}\simeq 10$ \emph{and} the average number of \emph{missing} 
links with an external community $(\langle m\rangle -\ell)\simeq 10$ 
where $\langle m\rangle$ is the average community size.
Thus, the communities remain well-defined particularly given their 
high edge density $p_{in}=1$. 

The two Potts models respond to the noise in the system in distinctly 
different ways in terms of how the community measure is calculated.
Noise complicates community assignment decisions for the RBCM because 
the configuration null model [the second term in \eqnref{eq:RBmodelCM}] 
incorporates the contribution of \emph{all} edges, including noise, 
for every node assignment evaluation even after a reliable solution 
``kernel'' is located during early stages of the solution dynamics 
(the metric effect of noise).

The APM evaluates all edges for community assignment decisions 
through relative energy calculations, 
but \figsref{fig:noisetest}{fig:noiseinittest} demonstrate that 
the best solution is often completely unaffected by the system 
noise if the algorithm can navigate sufficiently close to the 
solution.
Once an initial solution kernel evolves during the early stages 
of the algorithm dynamics, confusion caused by random system noise 
is often easily mitigated by the missing edge energy penalty
[the second term in \eqnref{eq:ourmodel}].
That is, the metric effect of noise on the APM is very favorable 
so that the main challenge caused by noise in the network is often 
due to the dynamical effect of noise (early incorrect assignments) 
that affects both models.

We could further improve the accuracy of the APM 
using a more robust, but much slower, algorithm such as SA.
Nevertheless, this benchmark illustrates that the energy landscape 
of the APM is more easily navigated, particularly for $\gamma =1$, 
than the RBCM. 
The energy landscape of the APM is more difficult to navigate 
for $\gamma\ll 1$ as compared to $\gamma\ge 1$ or 
when communities are not as strongly defined as they are in this test,
but the model maintains an exceptional accuracy 
\cite{ref:rzmultires,ref:lancLFRcompare}.

\section{Resolution Limit}  \label{sec:resolutionlimit}

The quantitative approaches of modularity \cite{ref:gn} and 
the RBPM \cite{ref:reichardt,ref:smcd} in \secref{sec:RBmodels} 
were implemented by incorporating global graph parameters 
into the models.
Both models marked important progress in the field of community 
detection, but Refs.\ \cite{ref:fortunato} and \cite{ref:kumpulaResLim} 
noted an unintended consequence of using global community measures 
--- an imposed resolution limit.
The resolution limit restricts the solutions of affected models 
so that they cannot correctly resolve all communities of a system 
in certain non-pathological cases.
For modularity and the RBCM, the number of communities in a 
system tends toward $\sqrt{L}$ \cite{ref:fortunato} and 
$\sqrt{\gammarb L}$ \cite{ref:kumpulaResLim}, respectively.
The models have difficulty properly resolving small communities 
in large systems and may incorrectly divide large communities.

We first discuss local versus global measures, and we then 
illustrate the resolution limit for the RB Potts models,
including modularity as a special case.
We also show that the APM is free of resolution-limit effects 
because it is a local measure of community structure.

\subsection{Local vs global measures} \label{sec:localglobal}

Including global dependences in quantitative community detection 
models was apparently rooted in the need to objectively determine 
the community structure of arbitrary graphs.
The assumption is that the global properties of the graph should 
imply its local community structure.
Global dependences make the partition solution objective since 
it allows the same quantitative model to automatically rescale 
to any graph, but they also became the central element that caused 
a resolution limit.

One suggested solution \cite{ref:fortunato,ref:kumpulaResLim}
to the resolution limit is to define a \emph{local} measure 
of community structure.
That is, community evaluations are made based only on local 
features of the graph in the neighborhood of the involved 
nodes and communities.
Some approaches that provide local community detection methods 
include clique percolation \cite{ref:palla,ref:kumpulacliqueperc},
analyzing random walks \cite{ref:rosvallmaprw,ref:chengshen},
``label propagation'' \cite{ref:LPA} (and one variant in \cite{ref:barberLPA}), 
and local variants of modularity \cite{ref:clausetlocal,ref:muff}.
See \secref{sec:weightedRBERlocal} and Appendix A regarding the RBER model.
The APM is also a strongly local measure of community structure.

Local models possess beneficial properties for solving some networks 
such as: 
large networks that are ``defined'' as the network is explored 
(e.g., the World Wide Web), 
incompletely known networks (e.g., social interactions),
coarse partitioning and refinement algorithms, 
and dynamic networks. 
Communities sufficiently isolated from graph changes do not 
have to be repetitively updated as the network is modified.
There exists work for modularity \cite{ref:modpreserve} that 
preserves the measure as the system is scaled, 
but a local model does not require this extra buttressing.
Further, several of the most accurate community detection methods 
\cite{ref:lancLFRcompare,ref:chengshen} are based on essentially 
local methods or models of community detection
\cite{ref:rosvallmaprw,ref:rzmultires,ref:chengshen}.

However, using a local measure of community structure returns 
to the subjectivity problem.
How does the model \emph{objectively} determine the community 
structure of arbitrary graphs?
Stated specifically for the APM, how does one choose the ``correct'' 
resolution(s) [\ie{}, value(s) of $\gamma$ in \eqnref{eq:ourmodel}] 
that will best solve the system?
Several answers to this problem are as follows although the concepts 
are not restricted to local models.

One approach is to define a community independent of the graph being solved.
For example, we might seek to identify all communities of ``close friends'' 
in a social network regardless of the size of the graph.
For the APM, \eqnref{eq:gammadensity} relates the model weight $\gamma$ 
to the minimum community edge density for all communities in a partition.

Some other methods, which are beyond the scope of this paper, define 
an algorithm or measure that can determine which resolutions 
(see \secref{sec:resolutiondef}) are the best partitions for the network.
Arenas \etal{} \cite{ref:multires} varied a weight parameter with modularity 
and tracked stable partitions.
Kumpula and co-workers \cite{ref:kumpulamultires,ref:heimo} 
as well as Fenn \etal{} \cite{ref:fenndynamic} also explored stability 
approaches for the RBCM.
Our multiresolution method \cite{ref:rzmultires} utilized information 
comparisons among independent solutions to quantitatively evaluate the 
best resolutions.
Zhang \etal{} \cite{ref:zhangsmall} used a topological weighting strategy.
Cheng and Shen \cite{ref:chengshen} used the stability of random walker 
diffusion dynamics to identify the most relevant resolutions.

\subsection{Circle of cliques}
\label{sec:circleofcliques}

\myfig{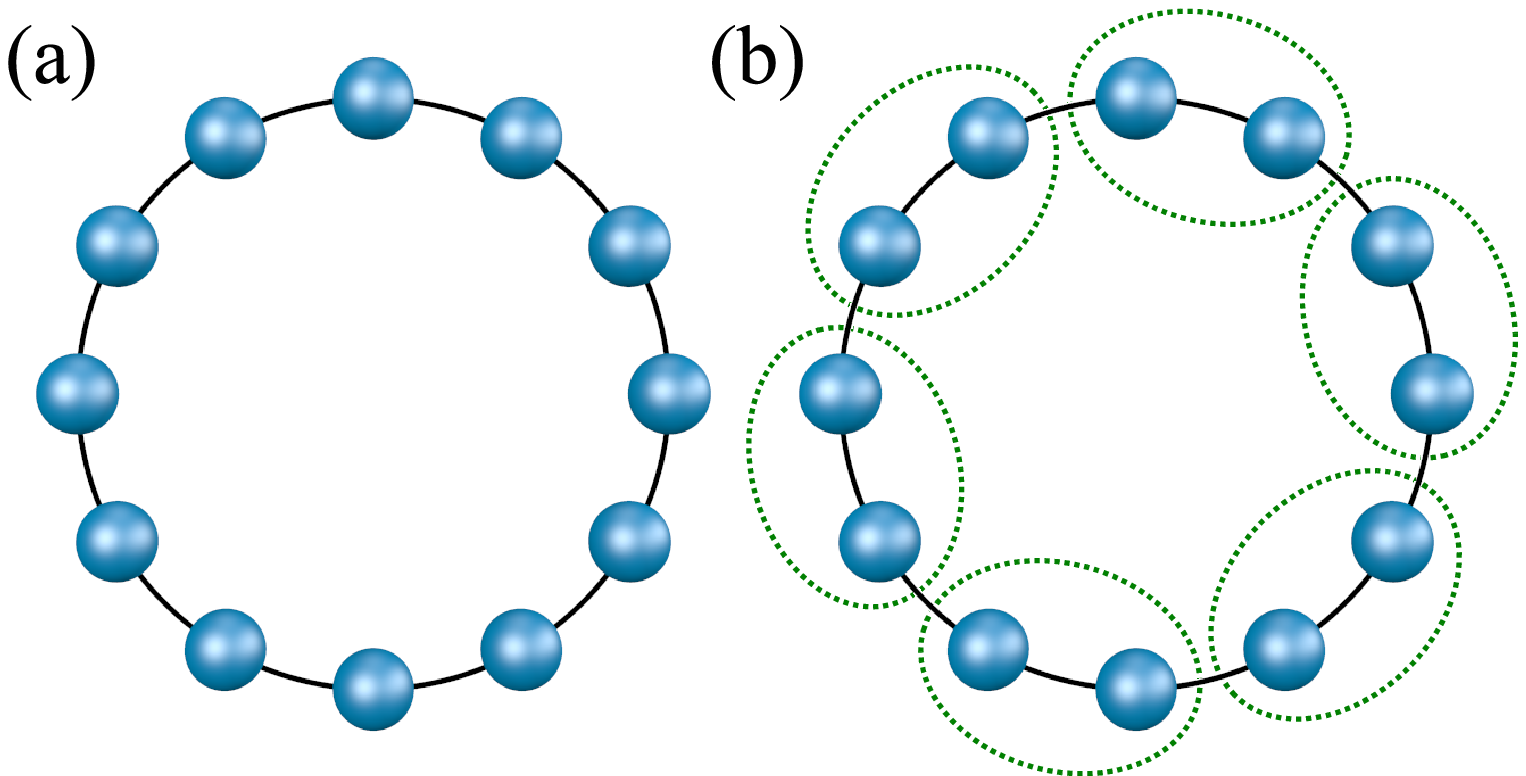}{(Color online) 
Each graph is a circle of cliques with $N$ total nodes and $L$ total edges.
(a)~A set of $q$ cliques with $m$ nodes each are connected in a circle 
by $q$ links.
(b)~$r$ consecutive cliques are each grouped together.
Intuitively, one would expect that any measure should resist 
merging these communities on any system scale (e.g., $N$, $L$, 
or $q$) if $m\ge 3$.
}{fig:circleofcliques}{0.85\linewidth}{t}

Fortunato and Barth{\'e}lemy \cite{ref:fortunato} and Kumpula 
\etal{} \cite{ref:kumpulaResLim} identified a resolution 
limit in the respective models in part by considering 
the unweighted system shown in \figref{fig:circleofcliques}, 
a set of $q$ cliques (maximally connected communities) 
connected in a circle by single edges.
In \subfigref{fig:circleofcliques}{a},
each clique is a separate community.
The total number of links is $L$ and the number 
of nodes in the system is $N$.
The total number of links between the cliques is $q$.
The number of nodes in each clique $m$ can be varied independently 
of $q$. 
From \eqnref{eq:ourmodel}, the APM energy is
\begin{equation}  
  \label{eq:circleHBa}
  E_\mathrm{a} = -\frac{1}{2}qm(m-1).
\end{equation}
This energy $E_\mathrm{a}$ has \emph{no finite extremum} 
with respect to \emph{any} global parameters of the graph.
The analogue to \eqnref{eq:circleHBa} for modularity and the RBCM
is where the resolution limit was demonstrated.
That is, those models have minima, $q^*_{mod} = \sqrt{L}$ 
and $q^*_{\phantom{}_{RB}} = \sqrt{\gammarb L}$, respectively,
in the expected number of communities.
Neither of these values correspond to the intuitive partition 
($q$ clique communities) for all system sizes. 
 
\subFigref{fig:circleofcliques}{b} depicts sets of $r$ cliques 
grouped together.
The specific conditions, based on $\gammarb$, for $r$ neighboring 
cliques to merge are given by the following relations.
The RBCM of \eqnref{eq:RBmodelCM}, using the configuration 
null model, includes modularity as a special case when $\gammarb =1$. 
Two neighboring cliques ($r=2$) \cite{ref:kumpulaResLim} will 
merge if
\begin{equation}  \label{eq:RBcirclegamma}
  \gammarb < \frac{q}{m(m-1) + 2}.
\end{equation}
The dependence on the number of cliques $q$ is a problem 
since this condition for $\gammarb$ can always be satisfied 
if $q$ is large enough (see also \secref{sec:mitigatedRL}).
For example, if $m=3$ and $\gammarb =1$, the cliques 
merge if $q>8$.

When using the RBER model with an \ermodel{} 
in \eqnref{eq:RBmodelER}, neighboring cliques merge if
\begin{equation}  \label{eq:RBcirclegammaER}
  \gammarb < \frac{q-1/m}{m(m-1) + 2}.
\end{equation}
We can always choose $q$ large enough to induce a merger 
of neighboring cliques for any $\gammarb$ (see also Appendix A).
These results generalize so that a resolution limit can 
be found to apply for an arbitrary choice of null model 
\cite{ref:kumpulaResLim} when using the RBPM
of \eqnref{eq:RBmodel}. The APM energy $E_\mathrm{b}$ of the configuration in 
\subfigref{fig:circleofcliques}{b} with $r$ merged cliques is 
\begin{eqnarray}    
  E_\mathrm{b} &=& -\frac{\gamma + 1}{2}qm(m-1) \nonumber \\
               & & \times\left[ 
            1 - \frac{\gamma}{\gamma + 1}\frac{rm-1}{m-1} 
              + \frac{2(r-1)}{rm(m-1)}  
          \right] .
  \label{eq:circleHB}
\end{eqnarray}
We compare the energies in \eqnarefs{eq:circleHBa}{eq:circleHB}
and find that $r=2$ cliques merge if 
\begin{equation}  \label{eq:circlegamma}
  \gamma < \frac{1}{m^2-1}.
\end{equation}
This merge condition depends \emph{only} on the \emph{local} 
variable $m$.
Solving the system with $\gamma = 1$ will ensure that 
neighboring cliques will not merge on \emph{any} global 
scale (e.g., $N$, $L$, or $q$) of the system.
Since $\gamma$ adjusts the weight applied to missing links, 
we can \emph{force} a merger of neighboring cliques if we 
reduce $\gamma$ to a sufficiently low value.
At $m=3$ for example, we can force a merger if $\gamma <1/8$.  

\subsection{Heterogeneous communities} 
\label{sec:heterogeneous}

\myfig{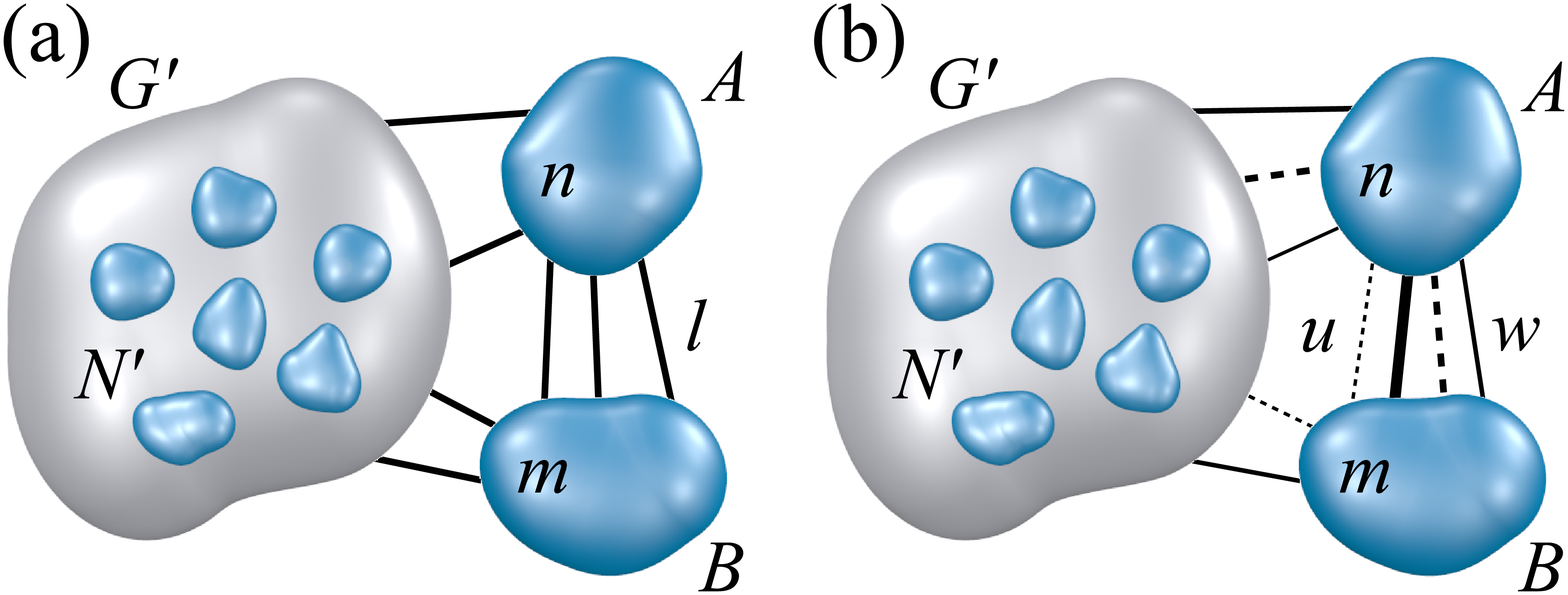}{(Color online) 
A large graph $G$ has $N$ nodes and three sub-divisions depicted, 
one potentially large sub-graph $G'$ with $N'$ nodes and two distinct 
communities $A$ and $B$ with $n$ and $m$ nodes, respectively.
In panel (a), $G$ is unweighted, and communities $A$ and 
$B$ are joined by $l$ edges.
In panel (b), $G$ is a weighted graph. 
For visualization purposes, solid lines depict weighted edges, 
dashed lines depict weighted missing links,
and the link thickness depicts a relative link weight.
Communities $A$ and $B$ are joined by weighted edges with a 
summed weight of $w$. 
Weighted \emph{missing} links have a summed weight of $u$.
All other graph features are left unspecified but are consistent 
with the community designations.
}{fig:heterograph}{0.9\linewidth}{t}

Resolution-limit effects can be exacerbated when communities 
of substantially different sizes are present.
Danon \etal{} addressed improvements for modularity 
to better resolve heterogeneous structures \cite{ref:danonhetero} 
with Newman's algorithm in \cite{ref:newmanfast}.
The APM deals with heterogeneous communities naturally. 

\subFigref{fig:heterograph}{a} depicts a large graph $G$ 
with three divisions.
Communities $A$ and $B$ have $n$ and $m$ nodes, respectively,
and are connected by $l$ edges.
Sub-graph $G'$ has $N'$ nodes with an unspecified structure.
For the RBPM of \eqnref{eq:RBmodel}, using a generic null model, 
the number of edges $l$ that causes communities $A$ and $B$ 
to merge is of the order \cite{ref:kumpulaResLim}
\begin{equation} \label{eq:RBmergeAB}
	l\gtrsim\frac{\gammarb nm}{N}.
\end{equation}
The RBER model yields a merge condition of 
\begin{equation} \label{eq:RBmergeABER}
	l > \frac{2L}{N(N-1)}\gammarb nm.
\end{equation}
In \eqnarefs{eq:RBmergeAB}{eq:RBmergeABER}, even for $l=1$ 
the merge conditions can be readily satisfied in large graphs 
for any reasonable value of $\gammarb$ due to the dependences 
on global graph parameters $L$ or $N$
(see also \secref{sec:mitigatedRL} and Appendix A). 

Our APM model merges communities $A$ and $B$ if
\begin{equation} \label{eq:ourmergeAB}
	l > \frac{\gamma}{\gamma + 1}nm.
\end{equation}
The merge condition is based only on $\gamma$ and the local 
community sizes $n$ and $m$.
For $\gamma=1$, even small communities merge with large ones 
only if there are many interconnections.  
The dependence on $\gamma$ is consistent with the purpose
of its introduction in \eqnref{eq:ourmodel} ---
to allow the model to vary the system resolution.

\subsection{Mitigated resolution limit}
\label{sec:mitigatedRL}

We return to the unweighted system of $q$ cliques 
in \figref{fig:circleofcliques} and \secref{sec:circleofcliques} 
to show that certain conditions will mitigate resolution-limit 
effects. 
By design, this circle of cliques was constructed to have an 
unambiguous intuitive answer.
Communities are not so clearly defined in practice, so we 
convert the cliques to communities with $\ell_{in}$ edges 
each, not necessarily maximally connected.  
We also increase the number of intercommunity edges 
so that each community has an average of $\ell_{out}$ edges connected 
to $s$ other communities ($qs\ell_{out}/2$ total external edges).
The original condition for the RBCM for neighboring cliques 
to merge [with $l_{in}=m(m-1)/2$, $l_{out}=1$, and $s=2$] is 
given by \eqnref{eq:RBcirclegamma}.
The new merge condition is
\begin{equation}  \label{eq:mitigatedRL}
  \gammarb <\frac{q\ell_{out}}{\left(2\ell_{in}+s\ell_{out}\right)}.
\end{equation}
High levels of noise [$s\simeq O(q)$ and $\ell_{out}\gtrsim O(1)$] 
tend to \emph{reduce} the effect of the resolution limit because 
the ratio is asymptotic to $\gammarb\simeq 1$. 

For the 
benchmark in \secref{sec:noisetest}, 
\eqnref{eq:mitigatedRL} explains how the RBCM can perform very well, 
despite a resolution limit, even when a large number of communities 
$q$ are present (we also subjectively evaluate many values of $\gammarb$).
On the other hand, more weakly defined communities [$\ell_{in}<m(m-1)/2$]
tend to increase resolution-limit effects, 
but system noise can substantially and positively influence the effects 
of the resolution limit.

\subsection{Locality of weighted Potts models} 
\label{sec:weightedlocal}

When considering weighted graphs, the introduction of (additional) 
global dependences should be a ``warning flag'' because global 
dependences are the source of the resolution limit. 
We show that the APM is remains a local model for weighted and 
directed graphs.

\subsubsection{Absolute Potts model}  
\label{sec:weightedAPM}

We generalize results from \secref{sec:heterogeneous} for the APM 
with an emphasis on weighted graphs including those with weighted 
missing links. 
Missing link weights correspond to levels of adversarial relations 
between nodes. ``Neutral'' relations use a weight $b_{ij}=1$
since a weight of $0$ is an inconsistent community detection 
model in general.
The following result also applies to directed graphs. 
Represented as a sum over communities, \eqnref{eq:ourmodel} becomes 
\begin{equation} \label{eq:ourmodelwqsum}
	\Ham_s (\{ \sigma \} ) = 
	  \sum_{s} \big( -w_s + \gamma ~\! u_s \big)  
\end{equation}
where $w_s$ and $u_s$ are the energy sums of \emph{connected} 
and \emph{missing} edges of community $s$, respectively.
For reference in \secref{sec:RBERwmodel}, 
the unweighted version of \eqnref{eq:ourmodelwqsum} is
\begin{equation} \label{eq:ourmodelqsum}
	\Ham_s (\{ \sigma \} ) = 
	  \sum_{s}{ \big[ -\left( \gamma+1 \right) l_s 
	                  + \gamma l^\mathrm{max}_s \big] }
\end{equation}
where $l_s$ is the number of edges and $l_s^\mathrm{max}$
is the maximum number of possible edges in community $s$.

In \subfigref{fig:heterograph}{b}, we use \eqnref{eq:ourmodelwqsum} 
to calculate the condition for two arbitrary communities $A$ and $B$ 
to merge in a general graph.
$A$ and $B$ are connected by edges with a total weight of $w$ and 
a total missing link weight of $u$.
The merge condition is almost trivially given by
\begin{equation} \label{eq:ourmergeABw}
	w>\gamma u.
\end{equation}
Note that this merge condition is based only on $\gamma$ and 
the connected or missing edges \emph{between} $A$ and $B$. 
The APM remains a local measure for general graphs in the 
strongest sense because node assignments are \emph{independent 
of the internal structure} of the communities 
(see also \secref{sec:weightedRBERlocal}).

\subsubsection{Weighted configuration RB Potts model}  
\label{sec:weightedRBCM}

A weighted generalization of the RBCM model is 
\begin{equation} \label{eq:RBmodelCMw}
  \Ham_{CM}^{w}(\{\sigma\}) = 
    \sum_{s} \left( -w_s+\gammarb\frac{W_s^2}{4W} \right) 
\end{equation}
where we express it as a sum over all communities.
$W$ is the total weight of all edges in the system and 
$W_s$ is the total weight of \emph{all} edges in community $s$
(including edges connected to \emph{other} communities).
As with the unweighted variant, this weighted model is 
necessarily already a global measure due to $W$ in the 
sum over $W_s^2$.

\subsubsection{Weighted \ErdosRenyi{} RB Potts model} 
\label{sec:RBERwmodel}

The weighted generalization of the RBER model of \eqnref{eq:RBmodelER} 
increases the global dependence of the model as it is proposed 
in \cite{ref:reichardt}.
We rewrite the \emph{unweighted} RBER model as a sum over communities 
\begin{equation} \label{eq:RBmodelERqsum}
	\Ham_{ER}(\{ \sigma \} ) = 
	  \sum_{s}{ \left( -l_s +\gammarb p~\! l^\mathrm{max}_s \right)}.
\end{equation}
\Eqnarefs{eq:ourmodelqsum}{eq:RBmodelERqsum} show that in the 
special but important case of \emph{unweighted} graphs, the APM 
and RBER models are coincidentally equivalent if we \emph{rescale} 
the null model weight by $\gammaer\equiv\gammarb p$ to explicitly 
remove the global density dependence (see Appendix A).  

We write a conceptual generalization of \eqnref{eq:RBmodelERqsum}
for weighted graphs which we use again in \secref{sec:weightedRBERlocal},
\begin{equation} \label{eq:RBmodelERwgeneral}
	\Ham_{ER}^{w}(\{ \sigma \} ) = 
	  \sum_{s}{ \left( -w_s + \gammarb p~\! w^\mathrm{max}_s \right)  }.
\end{equation}
Analogous to $l_s^\mathrm{max}$, $w_s^\mathrm{max}$ is 
the ``maximum weight sum'' of community $s$ which must be defined.
RB used one natural definition of 
($i$) $w_s^\mathrm{max}\equiv ~\!\overline{W} l_s^\mathrm{max}$
to obtain \cite{ref:reichardt} 
\begin{equation} \label{eq:RBmodelERwqsumglobal}
	\Ham_{ER}^{w}(\{ \sigma \} ) = 
	  \sum_{s}{\left( -w_s + \gammarb p\overline{W}~\! l_s^\mathrm{max}\right)}
\end{equation}
where $\overline{W}$ is the average weight over \emph{all} edges. 

In \subfigref{fig:heterograph}{b}, an arbitrary graph $G$ has three parts: 
two communities $A$ and $B$, and an arbitrary sub-graph $G'$.
Communities $A$ and $B$ are connected by a summed edge weight $w$. 
We ignore the missing link weight sum ($u=0$) since 
the model does not account for them.
Using \eqnref{eq:RBmodelERwqsumglobal}, the condition 
for the communities to merge is
\begin{equation} \label{eq:RBmergeABERw}
	  w > \gammarb p \overline{W} nm .
\end{equation}
The dependence on $\overline{W}$ allows arbitrary changes to 
independent parts of a graph to unintuitively affect each other.
For example, if we alter the edge weights in sub-graph $G'$, 
we change the average edge weight $\overline{W}$.
As a result, we indirectly change the condition for communities 
$A$ and $B$ to merge even though there are no local changes 
that affect $A$, $B$, or the links between them. 
This type of indirect effect caused by global parameters 
of the graph is at the heart of the resolution limit.

\subsubsection{Local ``\ErdosRenyi{}'' Potts model and ``weak'' locality}
\label{sec:weightedRBERlocal}

One can modify the weighted RBER model of \eqnref{eq:RBmodelERwgeneral} 
to create a local variant.
We briefly introduce our own variant since the comparison illustrates 
how ``strongly'' the APM defines a local measure of community structure.

Another natural interpretation of $w_s^\mathrm{max}$ 
in \eqnref{eq:RBmodelERwgeneral} is 
($ii$) $w_s^\mathrm{max}\equiv\overline{w}_s~\! l_s^\mathrm{max}$
where $\overline{w}_s$ is the average edge weight 
in the \emph{local} community $s$.
We also define $\gammaer\equiv\gammarb p$ to explicitly remove 
any dependence on the global density of the system.
(This was the initial form of the RBER model 
\cite{ref:reichardt}. See also Appendix A.)
In removing the density dependence $p$, the model is 
technically no longer an ``\ErdosRenyi{}'' Potts model; 
however, in this interpretation, 
\eqnref{eq:RBmodelERwgeneral} simplifies to
\begin{equation} \label{eq:RBmodelERws}
	\Ham_{ER}^{local}(\{ \sigma \} ) = 
	\sum_{s}{\wbar_s\left( -l_s+\gammaer ~\! l^\mathrm{max}_s\right)}.
\end{equation}
This variant uses almost the same energy sum as the \emph{unweighted} 
RBER model in \eqnref{eq:RBmodelER} except that total energy of community 
$s$ is weighted by $\overline{w}_s$. 
\Eqnref{eq:RBmodelERws} is a \emph{local} model in the sense that only 
parameters in the ``neighborhood'' of the local communities contribute 
to the energy, but it is a local model in a ``weaker'' sense than the 
APM because node assignments depend on the \emph{internal structure} 
(edge weights in this case) of the communities.

One can devise applications for such weakly local quality functions 
when influences within a graph need to be abstracted for efficiency 
or due to limited knowledge of the full details of the network 
(e.g., social networks with influential personalities).
However, despite being a local model, using these indirect dependences 
for community assignments (without associated edges between nodes) 
should elicit some caution because similar indirect 
effects on a global level are the source of the resolution 
limit for modularity and the RBPM.

\section{Examples} \label{sec:examples}



\myfig{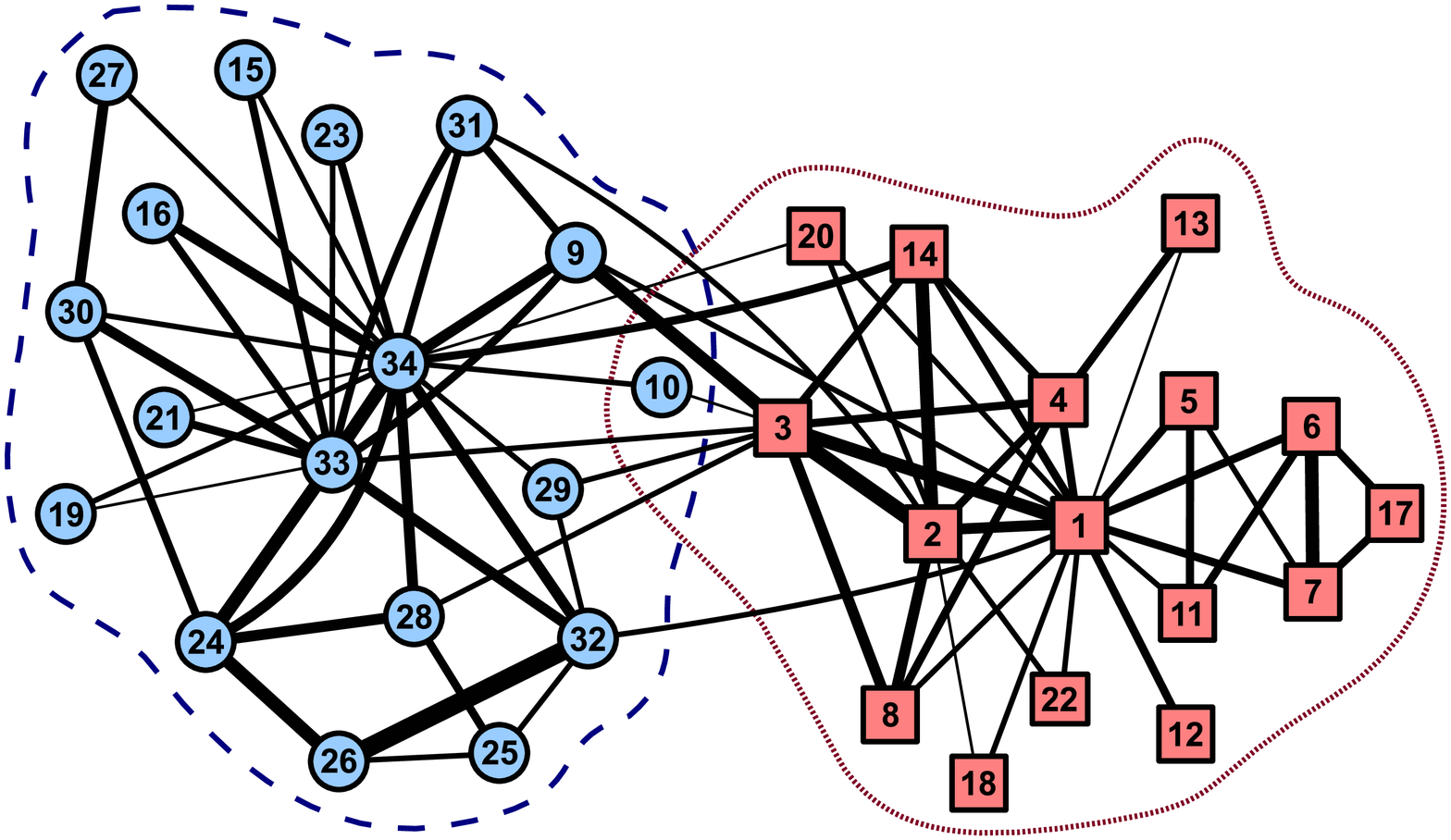}{(Color online) 
Graph depicts the Zachary karate club \cite{ref:zachary} solved 
with the APM using weighted edges (relative line thickness), 
$\gamma=1$, and $q=2$ communities by constraint.
All nodes except $10$ are correctly assigned.
By our analysis, this node appears frequently in both groups (see text).
}{fig:zacharyclub}{0.95\linewidth}{t!}

A common test is the Zachary karate club \cite{ref:zachary}.
It provides a small and real example of a social division that occurred 
while the group was under study.  
The graph consists of $34$ people with $78$ recognized relationships 
between them that are weighted according to the strength of the 
friendships (depicted by the relative line thickness).
We use the \emph{weighted} relations in  \eqnref{eq:ourmodel} with 
$b_{ij}=1$ and divide the graph into two parts by constraint as shown 
in \figref{fig:zacharyclub}.
Our algorithm correctly identifies the communities except for node $10$ 
which appears frequently in \emph{both} groups because there is no energy 
difference between the two assignments at $\gamma =1$.
(This is a rudimentary identification of an overlapping node using a method 
such as in \cite{ref:reichardt}.)
In the actual division, node $10$ associated with the group depicted 
by circles.
A more complete multiresolution analysis in \cite{ref:rzmultires} would 
correctly place node $10$.


We also construct a very large system similar to those defined 
in \secref{sec:noisetest}.
The system has $40\times 10^6$ nodes and
$L=1~\!157~\!634~\!899$ edges assigned in a power-law 
distribution of node degrees with an exponent $\alpha = -2$.
We specify the minimum and maximum degrees 
as $k_\mathrm{min}=20$ and $k_\mathrm{max}=500$, respectively.
The system is randomly partitioned into $q=2~\!443~\!782$ 
communities in a power-law distribution of community sizes 
with an exponent $\beta = -1$ with sizes ranging from 
$n_\mathrm{min}=10$ to $n_\mathrm{max}=25$ nodes.
The average internal community density is set to $p_{in}=0.95$.  
The average graph density is $p=1.45\times 10^{-6}$.

We solve the system with $\gamma =1/2$ in \eqnref{eq:ourmodel}
with the algorithm in \secref{sec:algorithm} using $t=1$ trial.  
(Random community edge assignments allow some nodes to be weakly 
connected to their intended communities. 
Using $\gamma =1/2$ ensures that all but the extreme outliers 
are properly assigned.)
The system was solved very accurately with $V=1.17\times 10^{-7}$ 
in $3.9$ hours on a single processor \cite{ref:computerusedone}.

\section{Conclusion} \label{sec:conclusion}

We present an exceptionally accurate and local spin-glass-type Potts 
model for community detection:
(1) our approach employs an absolute energy evaluation as opposed 
to a null model comparison.
(2) Its accuracy, even when using a greedy algorithm, is among 
the best of currently available algorithms.
(3) The model is robust to noise in the system.
(4) It is a local measure in the strongest sense of the term for 
unweighted, weighted (including weighted ``adversarial'' relationships), 
and directed graphs.
As such, it corrects a resolution-limit problem that affects other 
popular measures \cite{ref:gn,ref:fortunato,ref:smcd,ref:kumpulaResLim}.
(5) Heterogeneous community sizes are naturally resolved.
(6) The computational demand often scales as \bigO{tL^{1.3}} where 
$t$ is the number of optimization trials [generally \bigO{10} or less] 
and $L$ is the number of edges in the system. 
We have been able to accurately solve synthetic systems as large as 
$40\times 10^6$ nodes and over $10^9$ edges \cite{ref:computerusedone}.
In Ref.\ \cite{ref:rzmultires}, we illustrated in detail how this
core community detection method may be extended to systematically,
accurately, and rapidly identify general multiresolution structures.

\section*{ACKNOWLEDGMENTS}

We thank A. Lancichinetti and S. Fortunato for providing the SA code
and UCINet for network data made available on their website.
This work was supported by the LDRD DR on the physics of algorithms 
at LANL.

\appendix

\section{RESOLUTION LIMIT AND THE ERD\H OS-R\' ENYI POTTS MODEL}
\label{app:RBERnoreslimit}

For unweighted graphs, the RBER model of \eqnref{eq:RBmodelERqsum},
based on the \ermodel{}, is not inherently a global measure 
of community structure as is the RBCM, 
based on the configuration null model. 
The original model \cite{ref:reichardt} was defined without 
the density $p$ where $\gammaer\equiv \gammarb p$.
The \emph{ad hoc} inclusion of the density 
carried an implicit assumption that $\gammarb$ is constrained
to some range, perhaps by $\gammarb\simeq O(1)$ 
(otherwise, introducing a second constant is not meaningful). 
It then became a Potts model based on an \ermodel{}.

The justification for including the graph density in the model
was initially based on heuristic arguments about density 
inequalities that bounded the behavior of $\gammaer$.
Data were also presented using a common, but very small, 
benchmark (discussed in \secref{sec:GNtest}) 
that supported the approximation of $\gammaer\propto p$.
However, the approximation is not generally applicable.
For example, between the systems
in \sectref{sec:GNtest}{sec:APMvsRB}{sec:examples}, 
we would need to vary $\gammarb$ by at least $3$ orders 
of magnitude (and arbitrarily larger if we increase the system 
size in \secref{sec:noisetest}) if we wish to consistently 
identify the most accurate solution for each system.
If we remove the constant (but graph dependent) density $p$, 
we trivially remove from \eqnref{eq:RBmodelERqsum} 
any dependence on global graph parameters.

This change is more than a pedantic exercise.
Connecting the RBER model to the system density allowed it 
to automatically scale to solve arbitrary graphs in a semiobjective 
manner (see \secref{sec:localglobal}), 
but it also appeared to impose a resolution limit \cite{ref:kumpulaResLim}. 
Trivially removing the global dependence on $p$ effectively 
``eliminates'' the resolution limit for the model if one 
reinterprets the meaning of the original model weight $\gammaer$.
With this change, we assert that there is 
\emph{no genuine resolution limit} in the \emph{unweighted} 
RBER model as it was originally presented in \cite{ref:reichardt} 
without the density dependence $p$.

The second term in \eqnref{eq:RBmodelERqsum} indicates that $\gammaer$ 
specifies the fraction of $l_s^\mathrm{max}$ 
that each community must have before it has an energy less than zero.
Thus, we reinterpret $\gammaer$ 
as the minimum edge density of each community in a solved partition
(or the maximum external edge density \cite{ref:reichardt}), but this 
minimum density is enforced through only \emph{local} constraints.
The cost for this freedom is that we must \emph{choose} the ``correct'' 
weights $\gammaer$ for each graph, but the best choices are not arbitrary.

After removing $p$, we re-analyze the resolution-limit results 
obtained for the RBER model 
in \secsref{sec:circleofcliques}{sec:heterogeneous}. 
Using \subfigref{fig:heterograph}{a}, 
the original condition for two arbitrary unweighted communities 
$A$ and $B$ to merge is given by \eqnref{eq:RBmergeABER}.
Without $p$, the new merge condition is
\begin{equation}
	l>\gammaer nm
\end{equation}
which is based only on \emph{local} variables of communities $A$ and $B$ 
and the independently set $\gammaer$.

For the circle of cliques depicted in \figref{fig:circleofcliques}, 
the original merge condition is given by \eqnref{eq:RBcirclegammaER}.
The new condition for two neighboring cliques to merge is 
\begin{equation} \label{eq:RBmergenoRL}
  \gammaer < \frac{1}{m^2}.
\end{equation}
Using the reinterpretation of $\gammaer$, the value of $\gammaer =1/2$ 
demands at least a $50\%$ edge density for each community to be valid. 
At $m=3$, \eqnref{eq:RBmergenoRL} demands $\gammaer <1/9$ for a merger 
to occur. 
Therefore, at $\gammaer = 1/2$ the model will not experience a resolution 
limit effect for \emph{any} global scale of $N$, $L$, or $q$ for cliques 
of size $m\ge 3$.

After removing the density and reinterpreting $\gammaer$, 
the model is not genuinely subject to a resolution limit
because the constraints that define the community structure 
are enforced \emph{locally}.
We can then apply concepts mentioned in \secref{sec:localglobal}
to solve graphs with a local community measure.
One caveat is that the locality of the RBER model does not extend 
as naturally to weighted systems (see \secref{sec:weightedlocal}).

\section{VARIATION OF INFORMATION} 
\label{app:vi}

We use the variation of information metric \cite{ref:vi} 
for our accuracy tests in \secref{sec:APMvsRB}.
The probability that a randomly selected node from partition 
$A$ will be a member of community $k$ is $P(k) = n_k/N$, 
where $n_k$ is the number of nodes in community $k$
and $N$ is the total number of nodes in the system. 
The Shannon entropy is 
\begin{equation} \label{eq:Pkentropy}
  H(A) = -\sum_{k=1}^{q_A} \frac{n_k}{N}\log\frac{n_k}{N}
\end{equation}
where $q_A$ is the number of communities in partition $A$.

The mutual information $I(A,B)$ evaluates the level 
of interdependence in two sets of data.
We define a ``confusion matrix'' for partitions $A$ and $B$ 
by identifying how many nodes $n_{ij}$ of community $i$ 
of partition $A$ are in community $j$ of partition $B$. 
The mutual information is 
\begin{equation} \label{eq:mi}
  I(A,B) = 
  \sum_{i=1}^{q_A}\sum_{j=1}^{q_B} \frac{n_{ij}}{N} 
  \log\left(\frac{n_{ij} N}{n_i n_j}\right)  
\end{equation}
where $n_i$ is the number of nodes in community $i$ 
of partition $A$ and $n_j$ is the number of nodes 
in community $j$ of partition $B$.
The variation of information $V(A,B)$ is 
\begin{equation} \label{eq:vi} 
  V(A,B) = H(A) + H(B) - 2I(A,B),
\end{equation}
which measures the ``distance'' in information between 
the two partitions $A$ and $B$.
The range of values for VI is $0\le V(A,B)\le\log N$.
We use base $2$ logarithms.

\section{EXAMPLE NOISE TEST SOLUTION WITH THE RBCM}  
\label{app:RBBestExample}

\myfig{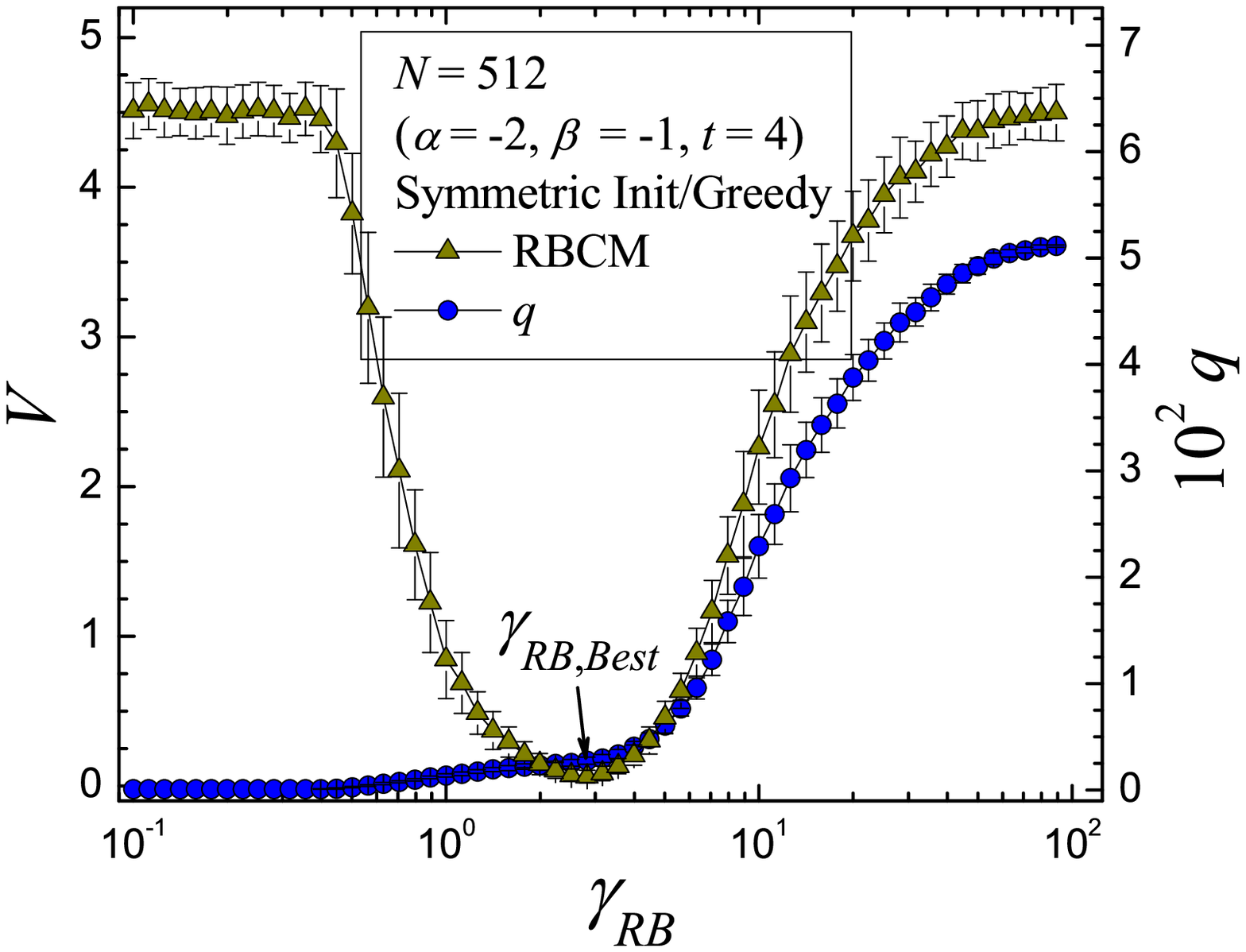}{(Color online) 
Plot of VI $V$ vs $\gammarb$ for the RBCM. 
In \secref{sec:noisetest}, we generate a set of strongly defined 
communities with varying levels of intercommunity noise $k_{out}$.
Using the greedy algorithm in \secref{sec:algorithm}, we compare 
the accuracy of solutions found with the RBCM to our APM.
Since the models operate differently, we compare the results 
in \figsref{fig:noisetest}{fig:noiseinittest} for our APM 
using $\gamma = 1$ to the \emph{best result} for the RBCM 
\emph{independently determined} for each $k_{out}$.
To this end, we increment $\gammarb$ by $20$ steps per decade 
and calculate VI for each solution using the \emph{known answer}.
We then select the best $\gammarb$ corresponding to the lowest VI 
average. 
This example is for a system with $N=512$ nodes (see \figref{fig:powersamplegraph}) 
with an average external degree $k_{out}\simeq 10$. 
See the text regarding other parameters defining the distribution 
of initial node degrees and community sizes.
Each point is an average over $100$ graphs.
}{fig:RBBestExample}{0.95\linewidth}{t}

In \secref{sec:noisetest}, we add noise to a strongly defined
system to test the accuracy of the RBCM of \eqnref{eq:RBmodelCM} 
compared to the APM of \eqnref{eq:ourmodel}.  
A sample system is depicted in \figref{fig:powersamplegraph},
and the accuracy results are summarized in 
\figsref{fig:noisetest}{fig:noiseinittest}.
For the APM, we solve the system with the model weight $\gamma=1$ 
for all graphs.
\Figref{fig:RBBestExample} shows an example of how we select the 
best result for the RBCM as compared to the known answer. 

We start with $\gammarb=0.1$ and geometrically increase the step 
size by $10^{1/20}$ (\ie{}, $20$ steps per decade of $\gammarb$).
This example is for $N=512$ nodes. 
The power-law distribution exponents are $\alpha =-2$ 
and $\beta =-1$ for the power-law degree and the community 
size distributions, respectively. 
Other parameters are: 
minimum and maximum community sizes $n_\mathrm{min}=4$ and 
$n_\mathrm{max}=50$, 
community edge densities $p_{in}=1$, average external degree 
(noise) $\kpowavg\simeq k_{out}\simeq 10$, maximum external degree 
$k_\mathrm{max}=100$, and $t=4$ trials per solution.
\Figref{fig:RBBestExample} shows only one \emph{best} answer 
for the RBCM. 
We average over $100$ graphs for each $\gammarb$, and the best 
VI average is plotted in \figsref{fig:noisetest}{fig:noiseinittest} 
with the result for the APM. 

\section{NOISE TEST ANALYSIS OF SA AT DIFFERENT STARTING TEMPERATURES}  
\label{app:SAtesting}

\myfig{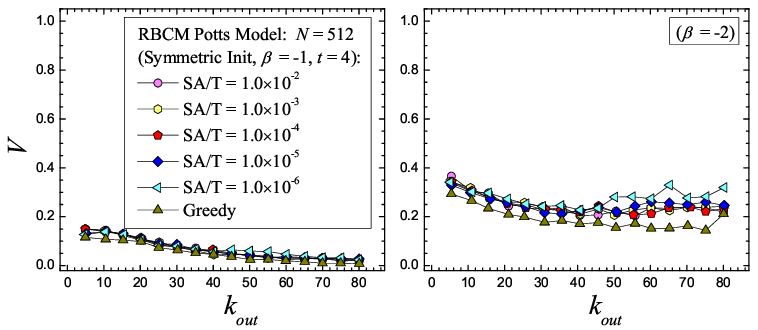}{(Color online)
Plot of VI $V$ vs the average external degree $k_{out}$ for the RBCM.  
In \secref{sec:noisetest}, we generate a set of strongly defined 
communities with high levels of intercommunity noise. 
In \figsref{fig:noisetest}{fig:noiseinittest}, we vary the initial 
average power-law degree $\kpowavg\simeq k_{out}$ and solve the 
networks using the greedy algorithm in \secref{sec:algorithm} and SA 
where we use a starting temperature of $T_0 = 1.0\times 10^{-4}$ 
for the $N=512$ node systems.
The greedy algorithm initialized into a symmetric initial state ($q_0=N$) 
\emph{outperforms} SA in accuracy when using the best $\gammarb$ 
(see Appendix C). 
In these plots, we further examine SA for a range of starting temperatures. 
Even with significantly higher starting temperatures, SA cannot exceed the 
accuracy of the greedy algorithm in this problem.
}{fig:RBSAtesting}{0.975\linewidth}{t!}

In \secref{sec:noisetest}, we construct a set of maximally connected 
communities with varied levels of intercommunity noise.
The systems are defined with a power-law distribution of community
sizes and an approximate power-law distribution of external noise 
(see \secref{sec:noisetest}).
We solve each system using the RBCM with the greedy algorithm 
in \secref{sec:algorithm} and SA both with $t=4$ trials.
In \figsref{fig:noisetest}{fig:noiseinittest}, we use starting temperatures 
of $T_0=1\times 10^{-4}$ for $N=512$ and $T_0=1\times 10^{-5}$ for $N=4096$.
Note that we scale the model energy by the number of edges in the 
system ($-1/L$) so that it is explicitly equivalent to the normalized 
modularity when $\gammarb =1$.
SA performs slightly \emph{worse} in accuracy than the greedy algorithm 
of \secref{sec:algorithm} using a symmetric initial state ($q_0 = N$).

Given this counter-intuitive result, in \figref{fig:RBSAtesting} 
we examine how the accuracy of the SA algorithm is affected by the 
algorithm's starting temperature.
We plot the average VI $V$ for the \emph{best} RBCM result (see Appendix C) 
versus the average external degree $k_{out}$.
We test starting temperatures spanning $5$ orders of magnitude for $N=512$ 
nodes. 
The cooling rate is fixed at $T_{i+1}=0.999 T_i$ where each 
step $i$ consists of $N$ randomly proposed state changes. 
For the highest temperatures, the computational time dramatically
increases due to a significantly longer cooling time with no 
significant improvement in accuracy.
Thus, a higher starting temperature for SA cannot improve its 
performance sufficiently to match the accuracy of the greedy 
algorithm in this problem. 

\section{EVIDENCE OF A PHASE TRANSITION IN COMMUNITY DETECTION}  
\label{app:phasetransition}

\myfig{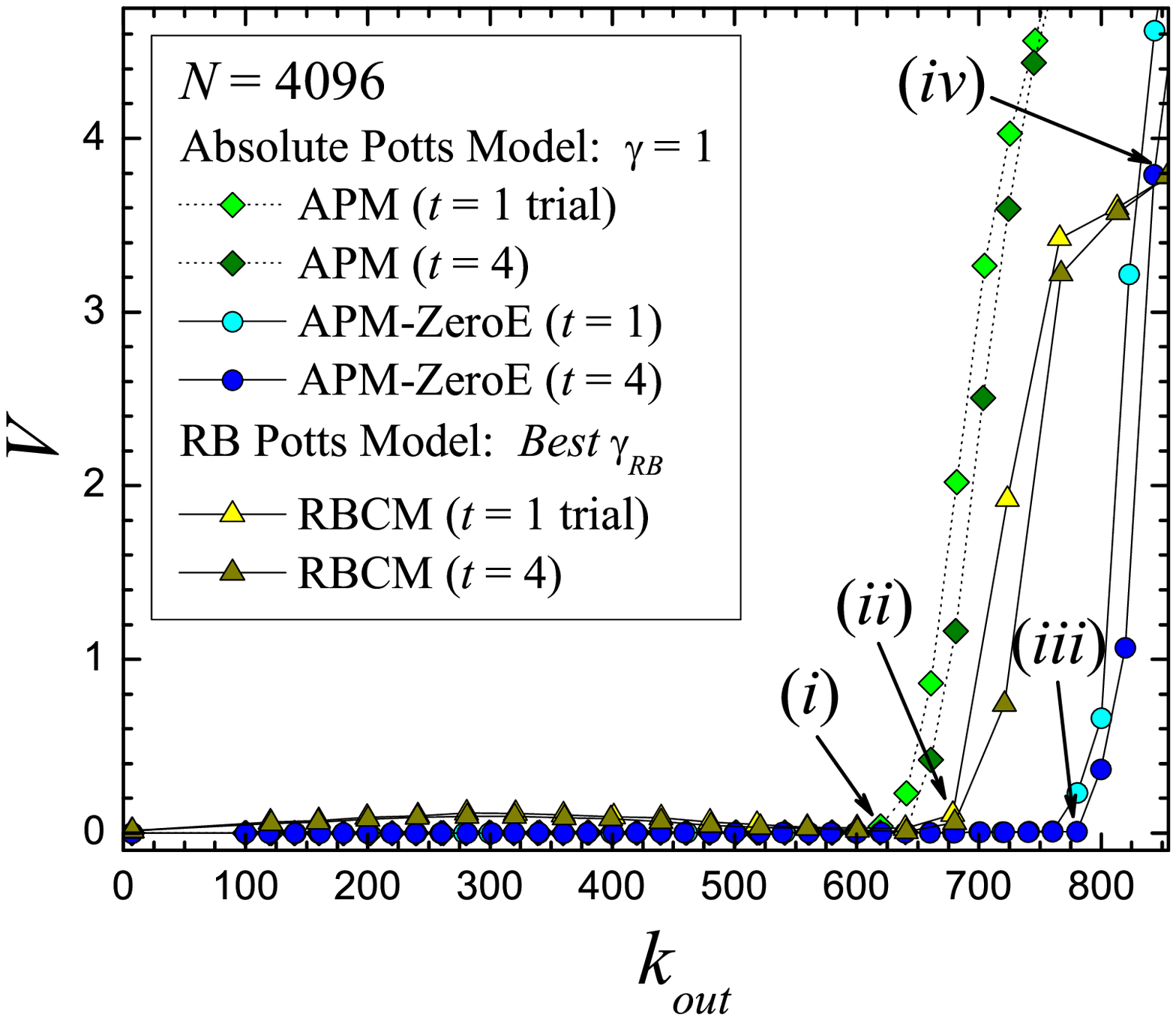}{(Color online)
Plot of VI $V$ vs the average external degree $k_{out}$ 
for the APM and RBCM models.
Similar to \secref{sec:noisetest}, we generate strongly 
defined communities with $N=4096$ nodes and high levels 
of intercommunity noise.
We use the greedy algorithm in \secref{sec:algorithm}
to solve the systems for both models.
The system is initially assigned a random power-law degree 
distribution with an exponent $\alpha =-2$, 
maximum degree $k_\mathrm{max}=1200$, and average degree 
$\kpowavg$ 
($k_{out}\simeq\kpowavg$).
Communities are assigned in a power-law distribution with 
an exponent $\beta =-1$, minimum size $n_\mathrm{min}=8$, 
maximum size $n_\mathrm{max}=24$, and density $p_{in}=1$.
The APM shows sharp accuracy transitions at ($i$) near $k_{out}\simeq 620$ 
(no zero energy moves) and at ($iii$) near $k_{out}\simeq 770$ 
(using zero energy moves).
These roughly correspond to a similar transition for the RBCM at ($ii$).  
See the text regarding ($iv$).
Each point is an average over $25$ graphs.
}{fig:phasetransition}{0.845\linewidth}{t}

In \figref{fig:phasetransition}, we construct a set of well-defined 
but noisy systems with $N=4096$ nodes from the benchmark 
in \secref{sec:noisetest}. 
An initial random degree distribution is assigned according to
a power law with an exponent of $\alpha = -2$, maximum degree 
$k_\mathrm{max}=1200$, and average degree $\kpowavg$.
Community sizes are assigned in a power-law distribution with an 
exponent of $\beta = -1$.
Minimum and maximum community sizes are $n_\mathrm{min}=8$ and 
$n_\mathrm{max}=24$, respectively.
We then maximally connect internal community edges (density $p_{in}=1$).  
We vary the average power-law degree $\kpowavg$ (the average external 
degree $k_{out}\simeq \kpowavg$) and solve the system with the APM and 
RBCM models using the algorithm in \secref{sec:algorithm}.

Features ($i$) and ($iii$) correspond to two related ``phase transitions'' 
of the system into a ``glassy'' state for our APM.
The ``complexity'' of the energy landscape dramatically increases 
near the ``critical points'' of $k_{out}^{{}_{(i)}} \simeq 630$ 
(where we \emph{disallow} zero energy moves) and 
$k_{out}^{{}_{(iii)}}\simeq 770$ 
(where we \emph{allow} zero energy moves), respectively.
After the transitions, our algorithm in \secref{sec:algorithm} 
is more easily trapped in a metastable state when navigating 
the energy landscape.
In the intermediate region, the APM can still \emph{almost perfectly} 
solve the system 
(in general problems, allowing zero energy moves does 
not usually result in such a drastic difference in accuracy).
A second aspect of these transitions (not depicted) is a generally rapid 
rise in the computational effort required to solve the system which peaks 
near the respective critical points.

Feature ($iii$) at $k_{out}^{{}_{(iii)}} \simeq 680$
shows that the RBCM displays a similar transition.
We speculate that the more complicated energy landscape of the RBCM
actually allows further optimization as compared to the APM 
when not utilizing zero energy moves in this problem.
At feature ($iv$), the best RBCM solution (see Appendix C) approaches 
a trivial partition with $q>3200$ communities.

This \emph{community detection transition} is similar to transitions 
in the k-SAT (k-SATisfiability) problem found by M{\'e}zard \etal{} 
\cite{ref:mezardpz}. 
The authors showed that the most difficult solutions for k-SAT problems 
are found along well-defined loci in the phase diagram of random 
satisfiability problems.
\Figref{fig:phasetransition} illustrates a similar transition 
in community detection.
We will address the phase-transition aspects of community detection 
in more detail in an upcoming publication.


\begin{thebibliography}{56}
\expandafter\ifx\csname natexlab\endcsname\relax\def\natexlab#1{#1}\fi
\expandafter\ifx\csname bibnamefont\endcsname\relax
  \def\bibnamefont#1{#1}\fi
\expandafter\ifx\csname bibfnamefont\endcsname\relax
  \def\bibfnamefont#1{#1}\fi
\expandafter\ifx\csname citenamefont\endcsname\relax
  \def\citenamefont#1{#1}\fi
\expandafter\ifx\csname url\endcsname\relax
  \def\url#1{\texttt{#1}}\fi
\expandafter\ifx\csname urlprefix\endcsname\relax\def\urlprefix{URL }\fi
\providecommand{\bibinfo}[2]{#2}
\providecommand{\eprint}[2][]{\url{#2}}

\bibitem[{\citenamefont{Danon et~al.}(2006)\citenamefont{Danon,
  D{\'i}az-Guilera, and Arenas}}]{ref:danonhetero}
\bibinfo{author}{\bibfnamefont{L.}~\bibnamefont{Danon}},
  \bibinfo{author}{\bibfnamefont{A.}~\bibnamefont{D{\'i}az-Guilera}},
  \bibnamefont{and} \bibinfo{author}{\bibfnamefont{A.}~\bibnamefont{Arenas}},
  \bibinfo{journal}{J. Stat. Mech.: Theory Exp.} 
  \bibinfo{year}{2006}, \bibinfo{pages}{P11010}.

\bibitem[{\citenamefont{Lancichinetti et~al.}(2009)\citenamefont{Lancichinetti,
  Fortunato, and Kert{\'e}sz}}]{ref:lanc}
\bibinfo{author}{\bibfnamefont{A.}~\bibnamefont{Lancichinetti}},
  \bibinfo{author}{\bibfnamefont{S.}~\bibnamefont{Fortunato}},
  \bibnamefont{and}
  \bibinfo{author}{\bibfnamefont{J.}~\bibnamefont{Kert{\'e}sz}},
  \bibinfo{journal}{New J. Phys.} \textbf{\bibinfo{volume}{11}},
  \bibinfo{pages}{033015} (\bibinfo{year}{2009}).

\bibitem[{\citenamefont{Girvan and Newman}(2002)}]{ref:gnsocbio}
\bibinfo{author}{\bibfnamefont{M.}~\bibnamefont{Girvan}} \bibnamefont{and}
  \bibinfo{author}{\bibfnamefont{M.~E.~J.} \bibnamefont{Newman}},
  \bibinfo{journal}{Proc. Natl. Acad. Sci. U.S.A.}
  \textbf{\bibinfo{volume}{99}}, \bibinfo{pages}{7821} (\bibinfo{year}{2002}).

\bibitem[{\citenamefont{Palla et~al.}(2005)\citenamefont{Palla, Der{\'e}nyi,
  Farkas, and Vicsek}}]{ref:palla}
\bibinfo{author}{\bibfnamefont{G.}~\bibnamefont{Palla}},
  \bibinfo{author}{\bibfnamefont{I.}~\bibnamefont{Der{\'e}nyi}},
  \bibinfo{author}{\bibfnamefont{I.}~\bibnamefont{Farkas}}, \bibnamefont{and}
  \bibinfo{author}{\bibfnamefont{T.}~\bibnamefont{Vicsek}},
  \bibinfo{journal}{Nature (London)} \textbf{\bibinfo{volume}{435}},
  \bibinfo{pages}{814} (\bibinfo{year}{2005}).

\bibitem[{\citenamefont{Clauset}(2005)}]{ref:clausetlocal}
\bibinfo{author}{\bibfnamefont{A.}~\bibnamefont{Clauset}},
  \bibinfo{journal}{Phys. Rev. E} \textbf{\bibinfo{volume}{72}},
  \bibinfo{pages}{026132} (\bibinfo{year}{2005}).

\bibitem[{\citenamefont{Blondel et~al.}(2008)\citenamefont{Blondel, Guillaume,
  Lambiotte, and Lefebvre}}]{ref:blondel}
\bibinfo{author}{\bibfnamefont{V.~D.} \bibnamefont{Blondel}},
  \bibinfo{author}{\bibfnamefont{J.-L.} \bibnamefont{Guillaume}},
  \bibinfo{author}{\bibfnamefont{R.}~\bibnamefont{Lambiotte}},
  \bibnamefont{and} \bibinfo{author}{\bibfnamefont{E.}~\bibnamefont{Lefebvre}},
  \bibinfo{journal}{J. Stat. Mech.: Theory Exp.} 
  \bibinfo{year}{2008}, \bibinfo{pages}{P10008}.

\bibitem[{\citenamefont{Xu and Chen}(2005)}]{ref:xuchen}
\bibinfo{author}{\bibfnamefont{J.~J.} \bibnamefont{Xu}} \bibnamefont{and}
  \bibinfo{author}{\bibfnamefont{H.}~\bibnamefont{Chen}}, \bibinfo{journal}{ACM
  Trans. Inf. Sys. Secur.} \textbf{\bibinfo{volume}{23}}, \bibinfo{pages}{201}
  (\bibinfo{year}{2005}).

\bibitem[{\citenamefont{Masuda}(2009)}]{ref:masudaimmune}
\bibinfo{author}{\bibfnamefont{N.}~\bibnamefont{Masuda}}, \bibinfo{journal}{New
  J. Phys.} \textbf{\bibinfo{volume}{11}}, \bibinfo{pages}{123018}
  (\bibinfo{year}{2009}).

\bibitem[{\citenamefont{Guimer{\`a} and Amaral}(2005)}]{ref:guimera}
\bibinfo{author}{\bibfnamefont{R.}~\bibnamefont{Guimer{\`a}}} \bibnamefont{and}
  \bibinfo{author}{\bibfnamefont{L.~A.~N.} \bibnamefont{Amaral}},
  \bibinfo{journal}{Nature (London)} \textbf{\bibinfo{volume}{433}},
  \bibinfo{pages}{895} (\bibinfo{year}{2005}).

\bibitem[{\citenamefont{Fortunato}(2010)}]{ref:fortunatophysrep}
\bibinfo{author}{\bibfnamefont{S.}~\bibnamefont{Fortunato}},
  \bibinfo{journal}{Phys. Rep.} \textbf{\bibinfo{volume}{486}},
  \bibinfo{pages}{75} (\bibinfo{year}{2010}).

\bibitem[{\citenamefont{Newman and Girvan}(2004)}]{ref:gn}
\bibinfo{author}{\bibfnamefont{M.~E.~J.} \bibnamefont{Newman}}
  \bibnamefont{and} \bibinfo{author}{\bibfnamefont{M.}~\bibnamefont{Girvan}},
  \bibinfo{journal}{Phys. Rev. E} \textbf{\bibinfo{volume}{69}},
  \bibinfo{pages}{026113} (\bibinfo{year}{2004}).

\bibitem[{\citenamefont{Blatt et~al.}(1996)\citenamefont{Blatt, Wiseman, and
  Domany}}]{ref:blatt}
\bibinfo{author}{\bibfnamefont{M.}~\bibnamefont{Blatt}},
  \bibinfo{author}{\bibfnamefont{S.}~\bibnamefont{Wiseman}}, \bibnamefont{and}
  \bibinfo{author}{\bibfnamefont{E.}~\bibnamefont{Domany}},
  \bibinfo{journal}{Phys. Rev. Lett.} \textbf{\bibinfo{volume}{76}},
  \bibinfo{pages}{3251} (\bibinfo{year}{1996}).

\bibitem[{\citenamefont{Reichardt and Bornholdt}(2004)}]{ref:reichardt}
\bibinfo{author}{\bibfnamefont{J.}~\bibnamefont{Reichardt}} \bibnamefont{and}
  \bibinfo{author}{\bibfnamefont{S.}~\bibnamefont{Bornholdt}},
  \bibinfo{journal}{Phys. Rev. Lett.} \textbf{\bibinfo{volume}{93}},
  \bibinfo{pages}{218701} (\bibinfo{year}{2004}).

\bibitem[{\citenamefont{Djijev}(2007)}]{ref:djidjev}
\bibinfo{author}{\bibfnamefont{H.~N.} \bibnamefont{Djijev}}, in
  \emph{\bibinfo{booktitle}{Algorithms and Models for the Web-Graph: Fourth
  International Workshop, WAW 2006, Revised Papers}}
  (\bibinfo{publisher}{Springer-Verlag}, \bibinfo{address}{Berlin, Heidelberg},
  \bibinfo{year}{2007}), Vol. \bibinfo{volume}{4936}, pp.
  \bibinfo{pages}{117--128}.

\bibitem[{\citenamefont{Reichardt and Bornholdt}(2006)}]{ref:smcd}
\bibinfo{author}{\bibfnamefont{J.}~\bibnamefont{Reichardt}} \bibnamefont{and}
  \bibinfo{author}{\bibfnamefont{S.}~\bibnamefont{Bornholdt}},
  \bibinfo{journal}{Phys. Rev. E} \textbf{\bibinfo{volume}{74}},
  \bibinfo{pages}{016110} (\bibinfo{year}{2006}).

\bibitem[{\citenamefont{Hastings}(2006)}]{ref:hastings}
\bibinfo{author}{\bibfnamefont{M.~B.} \bibnamefont{Hastings}},
  \bibinfo{journal}{Phys. Rev. E} \textbf{\bibinfo{volume}{74}},
  \bibinfo{pages}{035102(R)} (\bibinfo{year}{2006}).

\bibitem[{\citenamefont{Ispolatov et~al.}(2006)\citenamefont{Ispolatov, Mazo,
  and Yuryev}}]{ref:ispolatov}
\bibinfo{author}{\bibfnamefont{I.}~\bibnamefont{Ispolatov}},
  \bibinfo{author}{\bibfnamefont{I.}~\bibnamefont{Mazo}}, \bibnamefont{and}
  \bibinfo{author}{\bibfnamefont{A.}~\bibnamefont{Yuryev}},
  \bibinfo{journal}{J. Stat. Mech.: Theory Exp.} 
  \bibinfo{year}{2006}, \bibinfo{pages}{P09014}.

\bibitem[{\citenamefont{Fortunato and Barth{\'e}lemy}(2007)}]{ref:fortunato}
\bibinfo{author}{\bibfnamefont{S.}~\bibnamefont{Fortunato}} \bibnamefont{and}
  \bibinfo{author}{\bibfnamefont{M.}~\bibnamefont{Barth{\'e}lemy}},
  \bibinfo{journal}{Proc. Natl. Aca. Sci. U.S.A.}
  \textbf{\bibinfo{volume}{104}}, \bibinfo{pages}{36} (\bibinfo{year}{2007}).

\bibitem[{\citenamefont{Kumpula
  et~al.}(2007{\natexlab{a}})\citenamefont{Kumpula, Saram{\"a}ki, Kaski, and
  Kert{\'e}sz}}]{ref:kumpulaResLim}
\bibinfo{author}{\bibfnamefont{J.~M.} \bibnamefont{Kumpula}},
  \bibinfo{author}{\bibfnamefont{J.}~\bibnamefont{Saram{\"a}ki}},
  \bibinfo{author}{\bibfnamefont{K.}~\bibnamefont{Kaski}}, \bibnamefont{and}
  \bibinfo{author}{\bibfnamefont{J.}~\bibnamefont{Kert{\'e}sz}},
  \bibinfo{journal}{Euro. Phys. J. B} \textbf{\bibinfo{volume}{56}},
  \bibinfo{pages}{41} (\bibinfo{year}{2007}{\natexlab{a}}).

\bibitem[{\citenamefont{Ronhovde and Nussinov}(2008)}]{ref:rzone}
\bibinfo{author}{\bibfnamefont{P.}~\bibnamefont{Ronhovde}} \bibnamefont{and}
  \bibinfo{author}{\bibfnamefont{Z.}~\bibnamefont{Nussinov}},
  \bibinfo{journal}{e-print arXiv:0803.2548v1}. 

\bibitem[{ref({\natexlab{a}})}]{ref:hastingsnote}
\bibinfo{note}{The Potts model in~\cite{ref:hastings} also implicitly uses a
  version of a missing edge penalty, but the (accurate) implementation of the
  model is defined with a mean-field approximation.}

\bibitem[{\citenamefont{Ronhovde and Nussinov}(2009)}]{ref:rzmultires}
\bibinfo{author}{\bibfnamefont{P.}~\bibnamefont{Ronhovde}} \bibnamefont{and}
  \bibinfo{author}{\bibfnamefont{Z.}~\bibnamefont{Nussinov}},
  \bibinfo{journal}{Phys. Rev. E} \textbf{\bibinfo{volume}{80}},
  \bibinfo{pages}{016109} (\bibinfo{year}{2009}).

\bibitem[{\citenamefont{Traag and Bruggeman}(2009)}]{ref:traagPRE}
\bibinfo{author}{\bibfnamefont{V.~A.} \bibnamefont{Traag}} \bibnamefont{and}
  \bibinfo{author}{\bibfnamefont{J.}~\bibnamefont{Bruggeman}},
  \bibinfo{journal}{Phys. Rev. E} \textbf{\bibinfo{volume}{80}},
  \bibinfo{pages}{036115} (\bibinfo{year}{2009}).

\bibitem[{ref({\natexlab{b}})}]{ref:RBERnoisenote}
\bibinfo{note}{For \emph{unweighted} graphs, the RBER model should be
  equivalent to the APM if we ignore the density dependence $p$ representing
  the \ermodel{}. For example, on the benchmark in \secref{sec:noisetest}, the
  model would use $\gammaer\equiv\gammarb p = 1/2$ regardless of the system
  size or level of noise. See Appendix E for more discussion and
  \secsref{sec:weightedAPM}{sec:RBERwmodel} for differences between the models
  on weighted graphs.}

\bibitem[{\citenamefont{Danon et~al.}(2005)\citenamefont{Danon,
  D{\'i}az-Guilera, Duch, and Arenas}}]{ref:danon}
\bibinfo{author}{\bibfnamefont{L.}~\bibnamefont{Danon}},
  \bibinfo{author}{\bibfnamefont{A.}~\bibnamefont{D{\'i}az-Guilera}},
  \bibinfo{author}{\bibfnamefont{J.}~\bibnamefont{Duch}}, \bibnamefont{and}
  \bibinfo{author}{\bibfnamefont{A.}~\bibnamefont{Arenas}},
  \bibinfo{journal}{J. Stat. Mech.: Theory Exp.} 
  \bibinfo{year}{2005}, \bibinfo{pages}{P09008}.

\bibitem[{\citenamefont{Noack and Rotta}(2009)}]{ref:noack}
\bibinfo{author}{\bibfnamefont{A.}~\bibnamefont{Noack}} \bibnamefont{and}
  \bibinfo{author}{\bibfnamefont{R.}~\bibnamefont{Rotta}}, in
  \emph{\bibinfo{booktitle}{Experimental Algorithms}}, edited by
  \bibinfo{editor}{\bibfnamefont{J.}~\bibnamefont{Vahrenhold}}
  (\bibinfo{publisher}{Springer-Verlag Berlin, Heidelberg},
  \bibinfo{year}{2009}), Vol. \bibinfo{volume}{5526}, pp.
  \bibinfo{pages}{257--268}.

\bibitem[{\citenamefont{Lancichinetti and
  Fortunato}(2009)}]{ref:lancLFRcompare}
\bibinfo{author}{\bibfnamefont{A.}~\bibnamefont{Lancichinetti}}
  \bibnamefont{and}
  \bibinfo{author}{\bibfnamefont{S.}~\bibnamefont{Fortunato}},
  \bibinfo{journal}{Phys. Rev. E} \textbf{\bibinfo{volume}{80}},
  \bibinfo{pages}{056117} (\bibinfo{year}{2009}).

\bibitem[{\citenamefont{Clauset et~al.}(2004)\citenamefont{Clauset, Newman, and
  Moore}}]{ref:clausetlarge}
\bibinfo{author}{\bibfnamefont{A.}~\bibnamefont{Clauset}},
  \bibinfo{author}{\bibfnamefont{M.~E.~J.} \bibnamefont{Newman}},
  \bibnamefont{and} \bibinfo{author}{\bibfnamefont{C.}~\bibnamefont{Moore}},
  \bibinfo{journal}{Phys. Rev. E} \textbf{\bibinfo{volume}{70}},
  \bibinfo{pages}{066111} (\bibinfo{year}{2004}).

\bibitem[{\citenamefont{Raghavan et~al.}(2007)\citenamefont{Raghavan, Albert,
  and Kumara}}]{ref:LPA}
\bibinfo{author}{\bibfnamefont{U.~N.} \bibnamefont{Raghavan}},
  \bibinfo{author}{\bibfnamefont{R.}~\bibnamefont{Albert}}, \bibnamefont{and}
  \bibinfo{author}{\bibfnamefont{S.}~\bibnamefont{Kumara}},
  \bibinfo{journal}{Phys. Rev. E} \textbf{\bibinfo{volume}{76}},
  \bibinfo{pages}{036106} (\bibinfo{year}{2007}).

\bibitem[{\citenamefont{Gudkov et~al.}(2008)\citenamefont{Gudkov, Montealegre,
  Nussinov, and Nussinov}}]{ref:gudkov}
\bibinfo{author}{\bibfnamefont{V.}~\bibnamefont{Gudkov}},
  \bibinfo{author}{\bibfnamefont{V.}~\bibnamefont{Montealegre}},
  \bibinfo{author}{\bibfnamefont{S.}~\bibnamefont{Nussinov}}, \bibnamefont{and}
  \bibinfo{author}{\bibfnamefont{Z.}~\bibnamefont{Nussinov}},
  \bibinfo{journal}{Phys. Rev. E} \textbf{\bibinfo{volume}{78}},
  \bibinfo{pages}{016113} (\bibinfo{year}{2008}).

\bibitem[{ref({\natexlab{c}})}]{ref:computerusedone}
\bibinfo{note}{Systems were solved on AMD Opteron computers at $2.2$ -- $2.8$
  GHz with up to $48$ GB of random access memory.}

\bibitem[{\citenamefont{Newman}(2004)}]{ref:newmanfast}
\bibinfo{author}{\bibfnamefont{M.~E.~J.} \bibnamefont{Newman}},
  \bibinfo{journal}{Phys. Rev. E} \textbf{\bibinfo{volume}{69}},
  \bibinfo{pages}{066133} (\bibinfo{year}{2004}).

\bibitem[{\citenamefont{Boccaletti et~al.}(2007)\citenamefont{Boccaletti,
  Ivanchenko, Latora, Pluchino, and Rapisarda}}]{ref:boccaletti}
\bibinfo{author}{\bibfnamefont{S.}~\bibnamefont{Boccaletti}},
  \bibinfo{author}{\bibfnamefont{M.}~\bibnamefont{Ivanchenko}},
  \bibinfo{author}{\bibfnamefont{V.}~\bibnamefont{Latora}},
  \bibinfo{author}{\bibfnamefont{A.}~\bibnamefont{Pluchino}}, \bibnamefont{and}
  \bibinfo{author}{\bibfnamefont{A.}~\bibnamefont{Rapisarda}},
  \bibinfo{journal}{Phys. Rev. E} \textbf{\bibinfo{volume}{75}},
  \bibinfo{pages}{045102(R)} (\bibinfo{year}{2007}).

\bibitem[{ref({\natexlab{d}})}]{ref:RBCMGNtestnote}
\bibinfo{note}{In \figref{fig:accuracyplot}, $q$ is \emph{not constrained}
  during the solution dynamics for the RBCM/Symmetric data which results in a
  significantly lower accuracy than the data where $q=4$ is fixed. However, the
  model can achieve essentially the same accuracy for the unconstrained
  solution if we begin with a random state near $q_0=4$ communities. Work by
  Chauhan \etal{} \cite{ref:chauhanspectral} may indicate the expected $q$ in
  general problems, but it is uncertain whether this knowledge can be leveraged
  to improve the accuracy of a solution in general cases. In
  \secref{sec:initialconditions}, for example, we use a random initial state
  with $q_0\simeq q$, but the accuracy \emph{decreases} compared to a solution
  beginning with $q_0=N$.}

\bibitem[{ref({\natexlab{e}})}]{ref:datawebsite}
\bibinfo{note}{Data for the constructed system in Fig.\ \ref{fig:hierarchypic}
  can be found at http://physics.wustl.edu/zohar/communitydetection/}.

\bibitem[{ref({\natexlab{f}})}]{ref:logscalenote}
\bibinfo{note}{We use a log scale for the model weight $\gamma$ because of its
  relation to the minimum community edge density $p_\mathrm{min}$ in
  \eqnref{eq:gammadensity}. $\gamma$ identifies the targeted \emph{resolution}
  of the partition; and a log scale better captures, as compared to a linear
  scale, how the typical community density varies in the range of practical
  importance $0<\gamma\lesssim 19$ for unweighted graphs.}

\bibitem[{ref({\natexlab{g}})}]{ref:powerinitnote}
\bibinfo{note}{We generate a power-law degree distribution and randomly fill it
  in decreasing order of node degree left to fill. Given the high level of
  noise in our benchmark in \secref{sec:noisetest}, if one randomly selects
  successive pairs of nodes when adding new edges (such as is done in the PLOD
  algorithm~\cite{ref:plod}), nodes with smaller degrees will ``fill up'' first
  leaving nodes with larger degrees to be increasingly connected to each other
  as a group.}

\bibitem[{\citenamefont{Lancichinetti et~al.}(2008)\citenamefont{Lancichinetti,
  Fortunato, and Radicchi}}]{ref:lancbenchmark}
\bibinfo{author}{\bibfnamefont{A.}~\bibnamefont{Lancichinetti}},
  \bibinfo{author}{\bibfnamefont{S.}~\bibnamefont{Fortunato}},
  \bibnamefont{and} \bibinfo{author}{\bibfnamefont{F.}~\bibnamefont{Radicchi}},
  \bibinfo{journal}{Phys. Rev. E} \textbf{\bibinfo{volume}{78}},
  \bibinfo{pages}{046110} (\bibinfo{year}{2008}).

\bibitem[{\citenamefont{Good et~al.}(2009)\citenamefont{Good, de~Montjoye, and
  Clauset}}]{ref:goodMC}
\bibinfo{author}{\bibfnamefont{B.~H.} \bibnamefont{Good}},
  \bibinfo{author}{\bibfnamefont{Y.-A.} \bibnamefont{de~Montjoye}},
  \bibnamefont{and} \bibinfo{author}{\bibfnamefont{A.}~\bibnamefont{Clauset}},
  \bibinfo{journal}{Phys. Rev. E} \textbf{\bibinfo{volume}{81}},
  \bibinfo{pages}{046106} (\bibinfo{year}{2010}).

\bibitem[{\citenamefont{Radicchi et~al.}(2004)\citenamefont{Radicchi,
  Castellano, Cecconi, Loreto, and Parisi}}]{ref:radicchi}
\bibinfo{author}{\bibfnamefont{F.}~\bibnamefont{Radicchi}},
  \bibinfo{author}{\bibfnamefont{C.}~\bibnamefont{Castellano}},
  \bibinfo{author}{\bibfnamefont{F.}~\bibnamefont{Cecconi}},
  \bibinfo{author}{\bibfnamefont{V.}~\bibnamefont{Loreto}}, \bibnamefont{and}
  \bibinfo{author}{\bibfnamefont{D.}~\bibnamefont{Parisi}},
  \bibinfo{journal}{Proc. Natl. Acad. Sci. U.S.A.}
  \textbf{\bibinfo{volume}{101}}, \bibinfo{pages}{2658} (\bibinfo{year}{2004}).

\bibitem[{\citenamefont{Kumpula et~al.}(2008)\citenamefont{Kumpula, Kivel{\"a},
  Kaski, and Saram{\"a}ki}}]{ref:kumpulacliqueperc}
\bibinfo{author}{\bibfnamefont{J.~M.} \bibnamefont{Kumpula}},
  \bibinfo{author}{\bibfnamefont{M.}~\bibnamefont{Kivel{\"a}}},
  \bibinfo{author}{\bibfnamefont{K.}~\bibnamefont{Kaski}}, \bibnamefont{and}
  \bibinfo{author}{\bibfnamefont{J.}~\bibnamefont{Saram{\"a}ki}},
  \bibinfo{journal}{Phys. Rev. E} \textbf{\bibinfo{volume}{78}},
  \bibinfo{pages}{026109} (\bibinfo{year}{2008}).

\bibitem[{\citenamefont{Rosvall and Bergstrom}(2008)}]{ref:rosvallmaprw}
\bibinfo{author}{\bibfnamefont{M.}~\bibnamefont{Rosvall}} \bibnamefont{and}
  \bibinfo{author}{\bibfnamefont{C.~T.} \bibnamefont{Bergstrom}},
  \bibinfo{journal}{Proc. Natl. Acad. Sci. U.S.A.}
  \textbf{\bibinfo{volume}{105}}, \bibinfo{pages}{1118} (\bibinfo{year}{2008}).

\bibitem[{\citenamefont{Cheng and Shen}(2009)}]{ref:chengshen}
\bibinfo{author}{\bibfnamefont{X.-Q.} \bibnamefont{Cheng}} \bibnamefont{and}
  \bibinfo{author}{\bibfnamefont{H.-W.} \bibnamefont{Shen}},
  \bibinfo{journal}{e-print arXiv:0911.2308}. 

\bibitem[{\citenamefont{Barber and Clark}(2009)}]{ref:barberLPA}
\bibinfo{author}{\bibfnamefont{M.~J.} \bibnamefont{Barber}} \bibnamefont{and}
  \bibinfo{author}{\bibfnamefont{J.~W.} \bibnamefont{Clark}},
  \bibinfo{journal}{Phys. Rev. E} \textbf{\bibinfo{volume}{80}},
  \bibinfo{pages}{026129} (\bibinfo{year}{2009}).

\bibitem[{\citenamefont{Muff et~al.}(2005)\citenamefont{Muff, Rao, and
  Caflisch}}]{ref:muff}
\bibinfo{author}{\bibfnamefont{S.}~\bibnamefont{Muff}},
  \bibinfo{author}{\bibfnamefont{F.}~\bibnamefont{Rao}}, \bibnamefont{and}
  \bibinfo{author}{\bibfnamefont{A.}~\bibnamefont{Caflisch}},
  \bibinfo{journal}{Phys. Rev. E} \textbf{\bibinfo{volume}{72}},
  \bibinfo{pages}{056107} (\bibinfo{year}{2005}).

\bibitem[{\citenamefont{Arenas et~al.}(2007)\citenamefont{Arenas, Duch,
  Fern{\'a}ndez, and Gom{\'e}z}}]{ref:modpreserve}
\bibinfo{author}{\bibfnamefont{A.}~\bibnamefont{Arenas}},
  \bibinfo{author}{\bibfnamefont{J.}~\bibnamefont{Duch}},
  \bibinfo{author}{\bibfnamefont{A.}~\bibnamefont{Fern{\'a}ndez}},
  \bibnamefont{and}
  \bibinfo{author}{\bibfnamefont{S.}~\bibnamefont{Gom{\'e}z}},
  \bibinfo{journal}{New J. Phys.} \textbf{\bibinfo{volume}{9}},
  \bibinfo{pages}{176} (\bibinfo{year}{2007}).

\bibitem[{\citenamefont{Arenas et~al.}(2008)\citenamefont{Arenas,
  Fern{\'a}ndez, and G{\'o}mez}}]{ref:multires}
\bibinfo{author}{\bibfnamefont{A.}~\bibnamefont{Arenas}},
  \bibinfo{author}{\bibfnamefont{A.}~\bibnamefont{Fern{\'a}ndez}},
  \bibnamefont{and}
  \bibinfo{author}{\bibfnamefont{S.}~\bibnamefont{G{\'o}mez}},
  \bibinfo{journal}{New J. Phys.} \textbf{\bibinfo{volume}{10}},
  \bibinfo{pages}{053039} (\bibinfo{year}{2008}).

\bibitem[{\citenamefont{Kumpula
  et~al.}(2007{\natexlab{b}})\citenamefont{Kumpula, Saram{\"a}ki, Kaski, and
  Kert{\'e}sz}}]{ref:kumpulamultires}
\bibinfo{author}{\bibfnamefont{J.~M.} \bibnamefont{Kumpula}},
  \bibinfo{author}{\bibfnamefont{J.}~\bibnamefont{Saram{\"a}ki}},
  \bibinfo{author}{\bibfnamefont{K.}~\bibnamefont{Kaski}}, \bibnamefont{and}
  \bibinfo{author}{\bibfnamefont{J.}~\bibnamefont{Kert{\'e}sz}},
  \bibinfo{journal}{Fluct. Noise Lett.} \textbf{\bibinfo{volume}{7}},
  \bibinfo{pages}{L209} (\bibinfo{year}{2007}{\natexlab{b}}).

\bibitem[{\citenamefont{Heimo et~al.}(2008)\citenamefont{Heimo, Kumpula, Kaski,
  and Saram{\"a}ki}}]{ref:heimo}
\bibinfo{author}{\bibfnamefont{T.}~\bibnamefont{Heimo}},
  \bibinfo{author}{\bibfnamefont{J.~M.} \bibnamefont{Kumpula}},
  \bibinfo{author}{\bibfnamefont{K.}~\bibnamefont{Kaski}}, \bibnamefont{and}
  \bibinfo{author}{\bibfnamefont{J.}~\bibnamefont{Saram{\"a}ki}},
  \bibinfo{journal}{J. Stat. Mech.: Theory Exp.} 
  \bibinfo{year}{2008}, \bibinfo{pages}{P08007}.

\bibitem[{\citenamefont{Fenn et~al.}(2009)\citenamefont{Fenn, Porter, McDonald,
  Williams, Johnson, and Jones}}]{ref:fenndynamic}
\bibinfo{author}{\bibfnamefont{D.~J.} \bibnamefont{Fenn}},
  \bibinfo{author}{\bibfnamefont{M.~A.} \bibnamefont{Porter}},
  \bibinfo{author}{\bibfnamefont{M.}~\bibnamefont{McDonald}},
  \bibinfo{author}{\bibfnamefont{S.}~\bibnamefont{Williams}},
  \bibinfo{author}{\bibfnamefont{N.~F.} \bibnamefont{Johnson}},
  \bibnamefont{and} \bibinfo{author}{\bibfnamefont{N.~S.} \bibnamefont{Jones}},
  \bibinfo{journal}{Chaos} \textbf{\bibinfo{volume}{19}},
  \bibinfo{pages}{033119} (\bibinfo{year}{2009}).

\bibitem[{\citenamefont{Zhang et~al.}(2009)\citenamefont{Zhang, Zhang, ke~Xu,
  Tse, and Small}}]{ref:zhangsmall}
\bibinfo{author}{\bibfnamefont{J.}~\bibnamefont{Zhang}},
  \bibinfo{author}{\bibfnamefont{K.}~\bibnamefont{Zhang}},
  \bibinfo{author}{\bibfnamefont{X.}~\bibnamefont{ke~Xu}},
  \bibinfo{author}{\bibfnamefont{C.~K.} \bibnamefont{Tse}}, \bibnamefont{and}
  \bibinfo{author}{\bibfnamefont{M.}~\bibnamefont{Small}},
  \bibinfo{journal}{New J. Phys.} \textbf{\bibinfo{volume}{11}},
  \bibinfo{pages}{113003} (\bibinfo{year}{2009}).

\bibitem[{\citenamefont{Zachary}(1977)}]{ref:zachary}
\bibinfo{author}{\bibfnamefont{W.~W.} \bibnamefont{Zachary}},
  \bibinfo{journal}{J. Anthropol. Res.} \textbf{\bibinfo{volume}{33}},
  \bibinfo{pages}{452} (\bibinfo{year}{1977}).

\bibitem[{\citenamefont{Meil{\u a}}(2007)}]{ref:vi}
\bibinfo{author}{\bibfnamefont{M.}~\bibnamefont{Meil{\u a}}},
  \bibinfo{journal}{J. Multivariate Anal.} \textbf{\bibinfo{volume}{98}},
  \bibinfo{pages}{873} (\bibinfo{year}{2007}).

\bibitem[{\citenamefont{M{\'e}zard et~al.}(2002)\citenamefont{M{\'e}zard,
  Parisi, and Zecchina}}]{ref:mezardpz}
\bibinfo{author}{\bibfnamefont{M.}~\bibnamefont{M{\'e}zard}},
  \bibinfo{author}{\bibfnamefont{G.}~\bibnamefont{Parisi}}, \bibnamefont{and}
  \bibinfo{author}{\bibfnamefont{R.}~\bibnamefont{Zecchina}},
  \bibinfo{journal}{Science} \textbf{\bibinfo{volume}{297}},
  \bibinfo{pages}{812} (\bibinfo{year}{2002}).

\bibitem[{\citenamefont{Chauhan et~al.}(2009)\citenamefont{Chauhan, Girvan, and
  Ott}}]{ref:chauhanspectral}
\bibinfo{author}{\bibfnamefont{S.}~\bibnamefont{Chauhan}},
  \bibinfo{author}{\bibfnamefont{M.}~\bibnamefont{Girvan}}, \bibnamefont{and}
  \bibinfo{author}{\bibfnamefont{E.}~\bibnamefont{Ott}},
  \bibinfo{journal}{Phys. Rev. E} \textbf{\bibinfo{volume}{80}},
  \bibinfo{pages}{056114} (\bibinfo{year}{2009}).

\bibitem[{\citenamefont{Palmer and Steffan}(2000)}]{ref:plod}
\bibinfo{author}{\bibfnamefont{C.~R.} \bibnamefont{Palmer}} \bibnamefont{and}
  \bibinfo{author}{\bibfnamefont{J.~G.} \bibnamefont{Steffan}}, in
  \emph{\bibinfo{booktitle}{Global Telecommunications Conference, 2000.
  GLOBECOM '00}} (\bibinfo{publisher}{IEEE}, \bibinfo{address}{San Francisco,
  CA}, \bibinfo{year}{2000}), Vol.~\bibinfo{volume}{1}, pp.
  \bibinfo{pages}{434--438}.

\end{thebibliography}

\end{document}